\documentclass{article}

\usepackage[english]{babel}
\usepackage[utf8]{inputenc}
\usepackage[T1]{fontenc}
\usepackage{appendix}
\usepackage{amsmath}
\usepackage{bbm}
\usepackage{booktabs}
\usepackage{color}
\usepackage{hyperref}
\usepackage{braket}
\usepackage{algorithm}
\usepackage{algpseudocode}

\usepackage{pgfplots}
\usepackage{tikz}
\usetikzlibrary{shapes.geometric, arrows.meta, positioning}
\usetikzlibrary{decorations.pathreplacing,calc}

\usepackage[
    backend=biber,
    sorting=none,
    style=numeric-comp,
    maxnames=3
]{biblatex}

\usepackage[a4paper,top=3cm,bottom=2cm,left=3cm,right=3cm,marginparwidth=1.75cm]{geometry}

\usepackage{amsmath}
\usepackage{amsfonts}
\usepackage{graphicx}

\date{}

\title{Fast adaptive discontinuous basis  \\ sets for electronic structure}

\author{Yulong Pan\thanks{Department of Mathematics, University of California, Berkeley; Computational Research Division, Lawrence Berkeley National Laboratory (\texttt{yllpan@berkeley.edu}, \texttt{lindsey@berkeley.edu})} \and Michael
Lindsey\footnotemark[1]}

\begin{document}
\maketitle

\begin{abstract}
We develop a discontinuous Galerkin (DG) framework for automatically constructing adaptive basis sets for electronic structure calculations. By allowing basis functions to be discontinuous across element interfaces, our approach supports flexible combinations of atom-centered and polynomial basis sets, maintains favorable numerical conditioning, and induces structured sparsity of the one- and two-electron integrals, which we compute using specialised numerical integration strategies. In addition, we introduce a simple post-processing procedure to obtain continuous solutions if desired. We also introduce multigrid-preconditioned Poisson solvers that enable fast algorithms for both Hartree-Fock (HF) and density functional theory (DFT) calculations within our DG basis sets. Moreover, these basis sets naturally support adaptive multigrid preconditioning for the linear eigensolvers employed within the self-consistent field iteration for HF and DFT. Numerical experiments for HF and DFT demonstrate that our approach achieves chemical accuracy with modest basis sizes that compare favorably to the sizes of ordinary GTO basis sets achieving similar accuracy, while offering additional structured sparsity and improved computational scalability in the size-extensive limit. The framework thus provides a flexible route toward the construction of systematically improvable and structured adaptive basis sets for electronic structure theory.
\end{abstract}

\section{Introduction}
The formulation of many computational approaches to electronic structure theory begins with discretisation of the Hamiltonian. One usually chooses first a single-particle basis set, which induces an antisymmetric tensor product basis on the many-body Fock space. The single-particle basis set is typically constructed using atom-centred functions or planewaves.

Atom-centred basis sets, especially Gaussian-type orbitals (GTOs) \cite{werner2012molpro,fdez2019openmolcas,g16,neese2020orca,aidas2014d,saue2020dirac,apra2020nwchem,gordon2005advances,parrish2017psi4,sun2018pyscf,dovesi2014crystal14}, are widely used due to their capacity to represent nuclear cusps accurately with relatively few functions, while permitting analytical evaluation of the one- and two-electron integrals required for discretisation of the electronic structure Hamiltonian.
Alternatives to GTOs include, for example, numerical atomic orbitals \cite{blum2009ab}.

Atom-centered basis sets alone do not always yield a tractable pathway toward the complete basis set limit and can suffer from limited accuracy, particularly on metallic systems \cite{feller2018density,jensen2017elephant}. Furthermore, numerical conditioning can become a significant issue for large atom-centred basis sets, and the computational scaling with respect to the basis set size can be severe, especially for post-Hartree-Fock calculations.

In contrast, planewave bases offer a well-conditioned, systematically improvable, and highly structured basis set permitting fast algorithms and even (when smooth pseudopotentials are employed) spectral accuracy. However, especially for all-electron calculations, large basis set sizes are required to achieve chemical accuracy. Indeed, the planewave basis has uniform spatial resolution and as such is not adaptive to problem geometry, hence struggles to represent nuclear cusps.

Domain decomposition methods that combine atom-centred and planewave approaches have been developed to leverage the advantages of both perspectives \cite{blochl1994projector,madsen2001efficient,slater1937wave}. These divide the computational domain into (1) augmentation spheres near atomic nuclei, where atom-centred functions are used, and (2) interstitial regions discretised using planewaves. Such methods, however, require careful tuning of adjustable parameters such as sphere radii, and additional coupling conditions are needed to enforce regularity at sphere boundaries. Moreover, they are limited to periodic boundary conditions owing to their reliance on planewaves in the interstitial regions.

Beyond atom-centred and planewave approaches, a wide range of real-space discretisation methods have been considered, including finite differences \cite{chelikowsky1994finite, alemany2004real}, classical finite elements, \cite{tsuchida1995electronic,pask2005finite}, multiwavelets \cite{genovese2011daubechies,harrison2016madness,wind2022mrchem}, and enriched finite element basis sets
\cite{rufus2021fast, kanungo2023efficient,subramanian2025invdft}. The latter combine atom-centred enrichment functions with continuous piecewise polynomial basis sets. More broadly, many of these approaches employ adaptive strategies to conform basis sets to different geometries. However, to the best of our knowledge, these methods are not immediately extensible to the discretisation of the full electronic structure Hamiltonian in second quantisation, which requires both an orthogonal unconstrained single-particle basis set and a Galerkin projection of the Hamiltonian---including the Laplacian kinetic energy term.

Some recent directions in real-space discretisation have been proposed with a view toward structured and adaptive many-body calculations. These include pseudospectral diagonal basis sets \cite{white2017hybrid,white2019multisliced,white2023nested,lindsey2024fast} that allow fast structured treatment of the electron repulsion integrals (ERIs). A key limitation of these approaches is the difficulty of maintaining a compact basis set admitting fast operations, while accommodating arbitrary problem geometry.

In this paper, we allow for discontinuities in basis functions in order to introduce extra flexibility in the basis set construction. The presence of discontinuities is handled using the Discontinuous Galerkin (DG) framework, which has been widely applied to the numerical solution of partial differential equations, especially in the realm of computational fluid dynamics \cite{cockburn1998local,cockburn2012discontinuous,shu2013brief}. The DG method partitions a computational domain into non-overlapping domains, or elements. Discontinuities are permitted across elements, and basis functions are only required to be smooth within the confines of each element.

The advantage of allowing discontinuities is that basis functions can be constructed independently on individual elements. DG has been explored previously in the realm of electronic structure theory in several works \cite{lin2012adaptive,lin2017posteriori,lin2019convergence,hu2021high,hu20222}, which study a framework for building discontinuous adaptive basis functions for Kohn-Sham DFT. In their DGDFT framework, the authors show that very high accuracy can be obtained using only a small number of basis functions per atom.

However, the DGDFT method was not extended to the Hartree-Fock (HF) theory or to correlated many-body calculations. To the best of our understanding, a major reason for this gap is the lack of a fast solver for the Poisson equations required by real-space HF calculations. Furthermore, the DGDFT construction, like the enrichment finite element approach, relies on diagonalising local Hamiltonians for constructing local adaptive functions, rather than using established functions such as GTOs. While this preprocessing step can be parallelised, it nonetheless adds significant overhead to the calculation and necessitates additional design heuristics.

In this work, we introduce an alternative DG framework for discretising the electronic structure Hamiltonian. We construct basis sets adapted to the problem geometry that are built from primitive functions consisting possibly of both GTOs and polynomials, with supports suitably restricted to individual elements. A straightforward adaptive filtering procedure is applied to ensure orthogonality and to control the size of the resulting basis sets. Finally, to obtain fully continuous solutions using the discontinuous basis sets, we introduce an optional projection onto the continuous subspace embedded in the overarching discontinuous function space.

One perspective motivating our work is the goal of obtaining the same accuracy of a given highly accurate GTO basis set, without incurring the typical high computational cost of forming and storing all of the two-electron integrals. Indeed, we will see that our DG basis sets induced by a choice of an underlying GTO basis achieve accuracy at least as good and often significantly better than the ordinary GTO basis itself, while avoiding numerical conditioning woes that typically plague large GTO basis sets. Moreover, our DG basis enjoys structured sparsity and supports both both a fast multigrid Poisson solver and a fast adaptive multigrid \cite{pan2025geometric} preconditioner for the eigensolver appearing within the SCF for both HF and DFT calculations. Taken together these advantages yield an overall computational cost for mean-field calculations that scales almost linearly in the number of elements of the DG discretisation.

We contrast our Poisson solver with approaches based on fast multipole perspectives \cite{malhotra2015pvfmm,neese2025bubblepole}. Ultimately we adopt our new solver due to its superior empirical performance in our tests. Meanwhile, our adaptive multigrid preconditioner for the eigensolver should be contrasted with typical unpreconditioned approaches based on Chebyshev filtering \cite{banerjee2016chebyshev,kanungo2017large}. Observe that straightforward Fourier preconditioning based on the Laplacian contribution is unnatural because it requires the introduction of a grid of uniform resolution, and moreover, it is preferable to include the contribution of the nuclear potential in the preconditioner. To the best of our knowledge, the problem of designing preconditioners for the eigensolver remains relatively unexplored in previous work.

The paper is structured as follows. In Section \ref{sect:prelim}, we present the electronic structure problem, generalities about its discretisation, and mean-field formalisms (i.e., HF and DFT). Details of the Discontinuous Galerkin (DG) method are presented in Section \ref{sect:DG} and of our discontinuous basis set construction in Section \ref{sect:bases}. In Section \ref{sect:coulomb}, we outline the procedure for assembling the Coulomb integral terms in the Hamiltonian. In Section \ref{sect:auxiliary_mesh}, we provide details on the procedure for solving the relevant Poisson equations for both HF and DFT. The overall algorithm and its computational complexity are summarised in Section \ref{sect:alg_overview}, and numerical results are presented in Section \ref{sect:results}. Finally we describe some future directions and close in Section \ref{sect:conclusion}.

\section{Preliminaries} \label{sect:prelim}
\subsection{Overview}

We consider the electronic structure problem with $N_e$ electrons defined on $\mathbb{R}^3$. The Hamiltonian operator is given by
\begin{equation}
\label{eq:ham}
    -\sum_{i = 1}^{N_e} \Delta_i + \sum_{i=1}^{N_e} \sum_I \frac{Z_I}{|\mathbf{r}_i - \mathbf{R}_I|} - \sum_{1 \leq i < j \leq N_e} \frac{1}{|\mathbf{r}_i - \mathbf{r_j}|}, 
\end{equation}
acting on a space of antisymmetric $N_e$-particle wavefunctions. Electrons are indexed with lower-case $i$ and atomic nuclei with capital $I$, while $Z_I$ denotes the atomic number. The atom centres $\mathbf{R}_I$ are viewed as fixed and determine an external potential for the electrons.

Given some choice of orthonormal single-particle computational basis set $\{ \phi_i \, :\,  \mathbb{R}^3 \rightarrow \mathbb{R}, ~i=1,2,...,N_\phi \}$, the Galerkin projection of the Hamiltonian to the antisymmetric tensor product basis can be written in second-quantised format as 
\begin{equation} \label{eq:hamiltonian2}
    \sum_{i,j=1}^{N_\phi}(T_{ij}+U_{ij})  \, a_i^\dagger a_j + \sum_{i,j,k,l=1}^{N_\phi} V_{ijkl} \,  a_i^\dagger a_k^\dagger a_j a_l.
\end{equation}
Here $a_i$ (resp., $a_i^\dagger$) is the annihilation (resp., creation) operator for the $i$-th basis function. See \cite{lindsey2019quantum} for a mathematical introduction to second quantisation.

Meanwhile $T_{ij}, U_{ij}$ denote the one-electron kinetic and external potential contributions,  respectively: 
\begin{equation} \label{eq:TandU}
    T_{ij} = -\int_{\mathbb{R}^3} \phi_i (\mathbf{r}) \,  \Delta \phi_j (\mathbf{r}) ~d \mathbf{r}, \quad  U_{ij} = \sum_I \int_{\mathbb{R}^3} \phi_i(\mathbf{r}) \frac{Z_I}{|\mathbf{r}-\mathbf{R}_I|} \phi_j(\mathbf{r}) ~d \mathbf{r}, 
\end{equation}
and the tensor $V_{ijkl}$ of electron repulsion integrals (ERIs) is given by
\begin{equation} \label{eq:Vijkl}
    V_{ijkl} = \int_{\mathbb{R}^3} \int_{\mathbb{R}^3} \phi_i (\mathbf{r}_1) \phi_j (\mathbf{r}_1) \frac{1}{|\mathbf{r}_1-\mathbf{r}_2|} \phi_k (\mathbf{r}_2) \phi_l (\mathbf{r}_2) ~d\mathbf{r}_1 d\mathbf{r}_2.
\end{equation} 
Together these quantities fully specify the electronic structure Hamiltonian.

\subsection{Basis functions}
Single-particle basis functions are typically restricted to lie in the Sobolev space $H^1({\mathbb{R}^3})$, to ensure well-defined kinetic energy contributions involving the Laplace operator. However, this constraint narrows the function spaces available for representing single-particle states and can complicate the construction of localised or adaptive basis sets, particularly for complex molecular geometries.

In this work, we adopt an alternative formulation by relaxing this regularity requirement and instead consider basis functions in $L^2({\mathbb{R}^3})$. That is, the basis functions are allowed to be discontinuous, provided that these discontinuities occur along sets of measure zero. To handle such discontinuities while maintaining a consistent and coercive variational formulation, we employ the Discontinuous Galerkin (DG) framework, which enables the construction of weak derivative operators on discontinuous function spaces.

\subsection{Mean-field methods}
Effective single-particle methods treat the two-body electron-electron interactions at a mean-field level. This perspective reduces the full many-body Hamiltonian to an effective one-body operator, discretised using the single-particle basis. The resulting eigenvalue problem is solved to yield either molecular orbitals (MOs) in the case of the Hartree-Fock approximation or Kohn-Sham orbitals in the case of density functional theory.

\subsubsection{Hartree-Fock approximation} 
The Hartree-Fock (HF) approximation involves the solution of the nonlinear eigenvalue problem
\begin{equation} \label{eq:HF}
    \begin{split}
        \bigg( -\Delta + \sum_I \frac{1}{|\mathbf{r} - \mathbf{R}_I|} + V_\mathrm{H}[\rho] &+ V_\mathrm{X}[\{\psi_j \}] \bigg) \psi_k = \varepsilon_k \psi_k, \qquad k=1,\ldots,N_e \\
        \rho(\mathbf{r}) &= \sum_{k=1}^{N_e} | \psi_k(\mathbf{r}) |^2,
    \end{split}
\end{equation}
for the $N_e$ occupied MOs $\psi_1 , \ldots , \psi_{N_e} $, sorted in ascending order according to their associated energies $\varepsilon_k$. The HF approximation can be derived by optimization of the full many-body energy over the restricted class of Slater determinant wavefunctions.

Here $\rho(\mathbf{r})$ denotes the electron density constructed from the occupied orbitals. The Hartree potential $V_\mathrm{H}[\rho]$, which accounts for the classical electrostatic repulsion between electrons, is a diagonal operator given by the local expression
\begin{equation} \label{eq:hartree}
    V_\mathrm{H}[\rho](\mathbf{r},\mathbf{r}') = \delta(\mathbf{r}-\mathbf{r}')\int_{\mathbb{R}^3} \frac{\rho(\mathbf{r})}{|\mathbf{r} - \mathbf{r}'|} ~d\mathbf{r}'.
\end{equation}
which can be viewed as a functional of the electron density.

The Fock exchange operator $V_\mathrm{X}[\{\psi_j\}]$, a nonlocal operator arising from the antisymmetry constraint on the many-body wavefunction, is given by the integral kernel expression 
\begin{equation}\label{eq:exchange}
    V_\mathrm{X}[\{ \psi_j \}](\mathbf{r},\mathbf{r}') = -\sum_{k=1}^{N_e} \frac{\psi_k(\mathbf{r}) \psi_k(\mathbf{r}')}{|\mathbf{r}-\mathbf{r}'|}.
\end{equation}
Unlike the Hartree potential, the Fock exchange depends explicitly on the occupied orbitals and not solely on the electron density.

The nonlinear eigenvalue problem \eqref{eq:HF} is solved iteratively using a self-consistent field (SCF) iteration. In this work, we speed up the SCF convergence using the Adaptively Compressed Exchange (ACE) technique \cite{lin2016adaptively}, which involves an effective low-rank approximation to the Fock exchange $V_\mathrm{X}$ that is itself determined self-consistently. Additional details, including acceleration techniques, will be discussed in Section \ref{sect:scf}.

Adopting the bra-ket notation for inner products, the Hartree-Fock energy is calculated as 
\begin{equation}
    E_\mathrm{HF} = \sum_{i=1}^{N_e} \braket{\psi_i |T+U|\psi_i} + \frac{1}{2} \braket{\psi_i|V_\mathrm{H}[\rho] - V_\mathrm{X}[{\{\psi_j\}}]|\psi_i} + E_\mathrm{nuc}.
\end{equation}
Here, $T,U$ denote the kinetic energy and external operators corresponding to the matrices in \eqref{eq:hamiltonian2}, and $V_\mathrm{nuc}$ is the purely nuclear contribution to the energy
\begin{equation} \label{eq:Vnuc}
    E_\mathrm{nuc} = \frac{1}{2}\sum_{I \neq J}\frac{1}{|\mathbf{R}_I - \mathbf{R}_J|}.
\end{equation}

\subsubsection{Density functional theory}
Density functional theory (DFT) offers an alternative mean-field approach, in which the ground-state properties of an interacting many-electron system are determined from the electron density $\rho(\mathbf{r})$. The Kohn–Sham (KS) equations define a nonlinear eigenvalue problem
\begin{equation} \label{eq:KS}
    \begin{split}
    \bigg( -\Delta + \sum_I \frac{Z_I}{|\mathbf{r} - \mathbf{R}_I|} + V_\mathrm{H}[\rho]
      &+ V_\mathrm{XC}[\rho]
     \bigg) \psi_k = \varepsilon_k \psi_k, \qquad k=1,\ldots,N_e \\ 
     \rho(\mathbf{r}) &= \sum_{k=1}^{N_e} | \psi_k(\mathbf{r}) |^2,
    \end{split}
\end{equation}
for the Kohn–Sham orbitals $\psi_1, \ldots , \psi_{N_e} $. The nonlinear eigenvalue problem \eqref{eq:KS} is likewise solved iteratively via SCF iteration, as we detail in Section \ref{sect:scf}.

The Hartree potential is defined via \eqref{eq:hartree} as in the HF equations. The Fock exchange in DFT is replaced by the exchange-correlation potential $V_\mathrm{XC}$, which accounts jointly for the exchange and many-body correlation effects. The precise form of $V_\mathrm{XC}$ is unknown in general and approximated in practice; in this work we restrict our attention to the local density approximation (LDA), implemented in the Libxc library \cite{marques2012libxc}. More generally, so-called `hybrid functionals' such as B3LYP \cite{stephens1994ab} involving the exact exchange operator can offer better accuracy while incurring computational costs almost exactly the same as the Hartree-Fock approximation.

Again adopting bra-ket notation for inner products, the DFT energy is given by the expression
\begin{equation}
    E_\mathrm{DFT} = -\frac{1}{2} \sum_{i=1}^{N_e}\braket{\psi_i|T|\psi_i} + \int_{\mathbb{R}^3}  \rho(\mathbf{r})[U + \epsilon_\mathrm{XC}(\rho)] ~d\mathbf{r} + \frac{1}{2} \int_{\mathbb{R}^3} \int_{\mathbb{R}^3} \frac{\rho(\mathbf{r})\rho(\mathbf{r}')}{|\mathbf{r}-\mathbf{r}'|} ~d\mathbf{r}' d\mathbf{r} + E_\mathrm{nuc},
\end{equation}
where as in the HF case $T,U$ denote the kinetic energy and external potential operators, and $E_\mathrm{nuc}$ is the nuclear energy \eqref{eq:Vnuc}. The term $\epsilon_\mathrm{XC}$ is the exchange-correlation energy density corresponding to the potential $V_\mathrm{XC}$, which is also provided in Libxc.

\section{Discontinuous Galerkin framework} \label{sect:DG}

Now we provide the details of the Discontinuous Galerkin (DG) framework. In particular, we focus on the Symmetric Interior Penalty (SIP) method for discretising the Laplacian operator using discontinuous basis functions. Additionally, we discuss strategies for selecting the penalty parameter to ensure the stability and accuracy of the discretisation. Finally, we describe a simple post-processing procedure to project any discontinuous function discretised in our DG framework to the nearest continuous function.

\subsection{Preliminaries}
We employ the Discontinuous Galerkin framework to enable the use of piecewise continuous basis functions in $L^2({\mathbb{R}^3})$. We review some details of the DG framework pertinent to our application and refer the reader to \cite{hesthaven2008nodal} for a comprehensive treatment of the subject. 

To define a general DG discretisation, the domain $\mathbb{R}^3$ is first partitioned into a mesh consisting of non-overlapping elements $\mathcal{T} =\{K_m \, : \, m=1,\ldots, M\}$ such that $\bigcup_{m=1}^M K_m = \mathbb{R}^3$. The union of all element boundaries including boundary and interior faces is denoted by $\Gamma$. 

A broken function space $V_{\mathrm{DG}}$ is defined over this mesh as
\begin{equation}
    V_{\mathrm{DG}} = \{ v_h : v_h|_{K_m} \in V_m(K_m), \  m=1,\ldots, M \},
\end{equation}
where each $V_m(K_m)$ is a finite-dimensional subspace of $H^1(K_m)$, defined locally on the element $K_m$. Each element $K_m$ is associated with a local basis spanning $V_m(K_m)$, and the union of these local bases (extended by zero outside their respective elements) forms the global computational basis $\{ \phi_i, \ i=1,\ldots,N_\phi \}$ for $V_{\mathrm{DG}}$. 

In this work, we restrict our attention to basis functions that can be expressed as a linear combination of tensor product functions that admit the decomposition $f(x,y,z) = f_x(x)f_y(y)f_z(z)$.
Specific details regarding the basis functions used in this work are provided in Section~\ref{sect:bases}.

Any discontinuities of a function $f \in V_{\mathrm{DG}}$ are confined to the inter-element boundaries $\Gamma$. To treat such discontinuities, we introduce several standard definitions.

Consider a point $x_0 \in \Gamma$ lying on the shared interface between two elements $K_m, K_{m'}$. We define 
% the traces of $f$ approaching from within each element as
\begin{equation}
    f(x_0)^- := \lim_{x \rightarrow x_0,~ x \in K_m} f(x),\qquad  ~f(x_0)^+ := \lim_{x \rightarrow x_0,~ x \in K_{m'}} f(x).
\end{equation}
% where the superscripts $-$ and $+$ indicate values taken from the interior of $K_m, K_{m'}$, respectively.

The jump operator is then defined as
\begin{equation} \label{eq:jump_operator}
    [[ f ]] = f^- \mathbf{n}^- + f^+ \mathbf{n}^+,
\end{equation}
where $\mathbf{n}^-, \mathbf{n}^+$ are the outward-pointing unit normal vectors of $K_m, K_{m'}$, respectively, on the shared interface. 

Similarly, the average operator is defined by
\begin{equation}
    \{ f \} := \frac{1}{2}(f^- +f^+).
\end{equation}
Note that the jump operator produces a vector-valued quantity, while the average operator yields a scalar value.

\subsection{Mesh generation} \label{sect:meshing}
We restrict our focus in this work to molecular systems in free space. While the DG framework can easily accommodate periodic boundary conditions, we choose not to pursue a study of periodic systems in this work. Our aim is to establish and evaluate the method in the setting of isolated molecules, leaving extensions to solids for future work.

We refer to $\mathcal{T}$ as the orbital mesh hereafter. Orbitals are expected to exhibit sharp features near atomic centres $\mathbf{R}_I$ and to decay smoothly in regions far from the atoms. As such, finer mesh resolution is required near each $\mathbf{R}_I$, while coarser resolution suffices in the far field.

A common approach for constructing meshes with increased local resolution in three dimensions is through adaptive octree refinement. However, in this work we adopt a simpler strategy, since the inclusion of atom-centred basis functions has rendered any such refinement unnecessary in our empirical tests.

First we define a hexahedral bounding box of the form $[a_x,b_x] \times [a_y,b_y] \times  [a_z,b_z]$, where
\begin{equation} \label{eq:bounding_box}
    a_x = \min_I \{ R_{I,x} \} - B, \quad b_x = \max_I \{ R_{I,x} \} + B, 
\end{equation}
and similarly for the $y$- and $z$-dimensions. Here, $B$ is a buffer controlling the extent of the computational domain around the atoms. In practice, we set the buffer equal to one Bohr radius.

\begin{figure}
    \centering
    \includegraphics[scale=0.3]{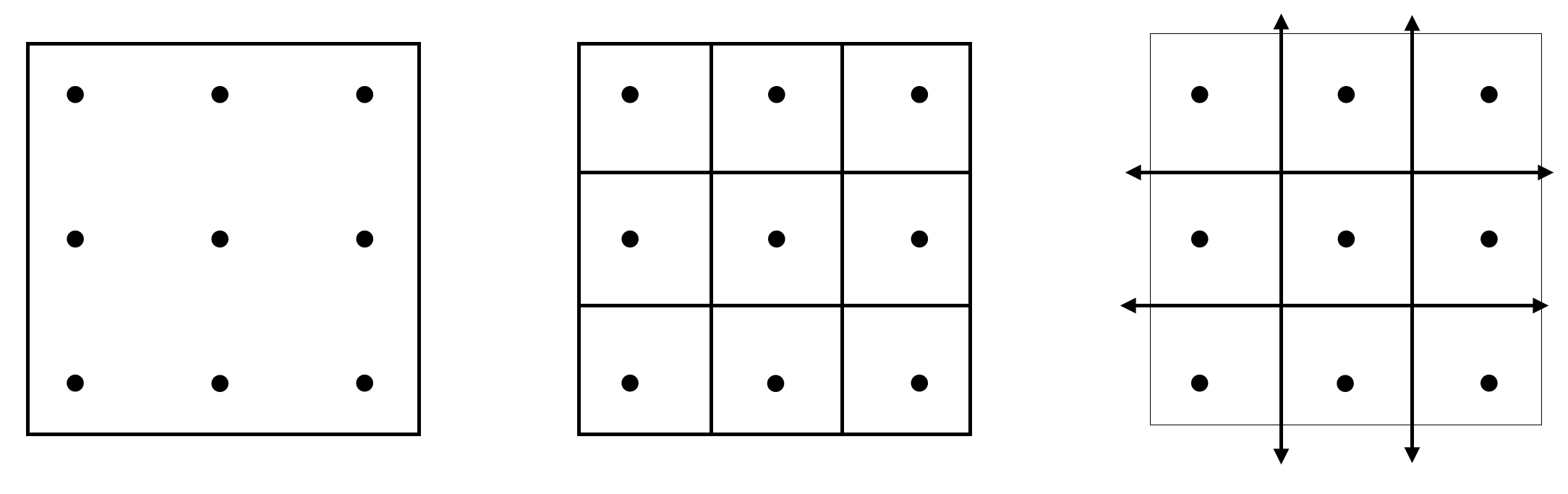}
    \caption{Schematic of free space mesh generation. A bounding box is placed around atom centres, shown in black, and carved up into uniform cells. Elements on domain boundaries are extended to infinity on sides with no neighbours.}
    \label{fig:mesh_free}
\end{figure}

The box is then partitioned into smaller uniform hexahedral elements subject to the constraint that each element contains at most $N_a$ atoms, to be specified by the user. To choose the partition, we select the integer tuple $(m_x, m_y, m_z)$ of elements per dimension minimising the sum $m_x m_y m_z$, subject to the constraint that at most $N_a$ atoms lie in each element. To account for the infinite support of the wavefunctions, elements located on boundary faces without immediate neighbours are extended to infinity. A schematic illustration of this mesh generation strategy in two dimensions is shown in Figure~\ref{fig:mesh_free}.

\subsection{Overlap matrix}
The DG overlap matrix $S$ is defined entrywise by 
\begin{equation} \label{eq:overlap}
    S_{ij} = \int_{\mathbb{R}^3} \phi_i(\mathbf{r}) \phi_j(\mathbf{r})~d\mathbf{r},
\end{equation}
analogously to continuous Galerkin discretisations. However, by contrast, the DG overlap matrix $S$ is block diagonal. This structure arises because each basis function has support only on a single element $K_m$. As a result, the entry $S_{ij} \neq 0$ if and only if any two basis functions $\phi_i,\phi_j$ are supported on the same element $K_m$. 

In our construction, we will ultimately impose that basis functions are pairwise orthogonal with respect to the standard $L^2$  inner product, yielding $S=I$. The procedure used to construct orthogonal basis functions is described in more detail in Section~\ref{sect:adaptive_truncation}.
%\ML{Aren't we going to orthogonalize our basis? It would be good to mention here that we can assume the overlap is identity}

\subsection{Laplacian matrix}

As we are focused only on infinite domains, we omit any boundary terms in the Laplacian weak form and focus only on inter-element interfaces of $\Gamma$.

A key challenge in defining a valid DG discretisation of the Laplacian operator arises from the discontinuities of functions in $V_{\mathrm{DG}}$. To address this difficulty, DG restricts volume integrals to the interiors of elements and introduces interface terms on $\Gamma$ to ensure consistency and stability.

% The bilinear form \ML{do we need form? or just matrix? you never define the matrix carefully} for the Laplacian takes the general form
% \begin{equation}
%     T(u,v) = \sum_{m=1}^M \int_{K_m} \nabla u(\mathbf{r}) \cdot \nabla v(\mathbf{r}) ~d\mathbf{r} + \int_\Gamma P(u,v) ~ds,
% \end{equation}
% \ML{the meaning of the second term is not clear... what is s?} where $P(u,v)$ is a penalty satisfying \ML{This is awkward.. maybe skip to the choice you make}
% \begin{equation}
%     P(u,u) = \begin{cases}
%         =0, ~u \in H^1,~ \\
%         >0, ~u \in L^2 \setminus H^1.~
%     \end{cases}
% \end{equation}
% This penalty suppresses non-physical modes introduced by discontinuities across elements and enforces weak continuity across interfaces.

In this work, we adopt the Symmetric Interior Penalty Discontinuous Galerkin (SIPDG) formulation \cite{arnold1982interior}, in which the discrete Laplacian matrix has entries given by
\begin{equation} \label{eq:sip_bilinear}
    T_{ij} = \sum_{m=1}^M \int_{K_m} \nabla \phi_i(\mathbf{r}) \cdot \nabla \phi_j(\mathbf{r}) ~d\mathbf{r} + \int_\Gamma -[[\phi_i]] \cdot \{ \nabla \phi_j \} - \{ \nabla \phi_i \} \cdot [[\phi_j]] + \sigma[[\phi_i]] \cdot [[\phi_j]] ~ds,
\end{equation}
where $\sigma > 0$ is a penalty parameter to be defined. This penalty term suppresses non-physical modes arising from discontinuities across element interfaces and enforces weak continuity.

The SIPDG formulation generalises the standard Laplacian bilinear form: all interface terms vanish in the case where the basis functions $\phi_i$ are globally continuous. It is well-established that this construction yields a discretisation that is both consistent and stable, provided $\sigma$ is chosen sufficiently large. For further theoretical details and alternative DG formulations, we refer the reader to \cite{arnold2002unified}.

\subsection{Penalty parameter} \label{sect:penalty}

We describe a procedure to choose the penalty parameter $\sigma$ in \eqref{eq:sip_bilinear}. Since this work focuses on tensor product basis function spaces on hexahedral elements for which the Laplacian is represented via a Kronecker sum \cite{pazner2018approximate}, it suffices to consider the one-dimensional Laplacian. We will explain below in more detail how the penalty is recovered in the multidimensional case.

For the one-dimensional case, we consider two neighbouring elements $K^- = [x_{k-1}, x_k]$ and $K^+ = [x_k, x_{k+1}]$ with local orthonormal basis functions $\phi_i^-$, $\phi_i^+$ respectively. Following \cite{arnold1982interior}, coercivity of the SIPDG bilinear form is ensured if $\sigma$ is chosen on the shared interface between the two elements to satisfy
\begin{equation}
    \sigma \geq \frac{C^2}{\epsilon h}, \quad h = \min( x_k - x_{k-1}, x_{k+1} - x_k ),
\end{equation}
where $C$ is a constant dependent on the basis, $h$ the characteristic length, and $\epsilon \in (0,1]$ a hyperparameter that can be tuned to adjust the size of the penalty beyond the minimal requirement for coercivity. In practice, we choose $\epsilon = \frac{1}{8}$ based on empirical performance. %\ML{This doesn't make sense really, the criterion is that you are bigger than something, so maybe you mean = where $\epsilon \leq 1$ as a hyperparameter controlling the factor}
%\ML{This is better. But you can still improve it. It feels like an open-ended journey} 

More specifically, the constant $C$ must be taken sufficiently large to ensure that the trace inequality 
\begin{equation}
\label{eq:Cprop}
    \left| \frac{d}{dx} v^\pm(x_j) \right|^2 \leq C^2 \bigg\| \frac{d}{dx} v(x) \bigg\|^2_{L^2 (K^\pm)}
\end{equation}
holds for arbitrary $v$ in the span of the basis functions $\phi_i^-$, $\phi_i^+$. Elementary computations in this orthonormal basis \cite{warburton2003constants} reveal that we can take 
\begin{equation} \label{eq:Csquared}
    C^2 := \max\left\{ \sum_i \left[ \frac{d}{dx} \phi_i^- (x_k) \right]^2, \sum_i \left[ \frac{d}{dx} \phi_i^+ (x_k) \right]^2 \right\}.
\end{equation}
For polynomial bases of degree at most $p$, an explicit formula is provided in \cite{warburton2003constants}
\begin{equation} \label{eq:pen_poly}
    C^2 = \frac{(p+1)^2}{h},
\end{equation}
while for general basis sets, $C$ can be computed numerically by evaluating \eqref{eq:Csquared}.

For higher-dimensional problems with tensor product basis functions on hexahedral elements, the penalty parameter can be determined by reducing to the one-dimensional setting. This simplification is possible because, on each interface between two neighboring elements, the boundary contributions from the SIPDG formulation are governed by terms that act only in the direction normal to the interface, which are aligned with the standard Cartesian basis vectors. As a result, the penalty parameter in high dimensions can taken as the same penalty from the corresponding one-dimensional problem along that normal direction, illustrated in Figure \ref{fig:pen3d}.

\begin{figure}[h]
    \centering
    \begin{tikzpicture}[scale=1]

    % Cube size
    \def\L{2}

    % === Left Cube ===
    \draw[thick] (0,0,0) -- (\L,0,0) -- (\L,\L,0) -- (0,\L,0) -- cycle;
    \draw[thick] (0,0,\L) -- (\L,0,\L) -- (\L,\L,\L) -- (0,\L,\L) -- cycle;
    \draw[thick] (0,0,0) -- (0,0,\L);
    \draw[thick] (\L,0,0) -- (\L,0,\L);
    \draw[thick] (\L,\L,0) -- (\L,\L,\L);
    \draw[thick] (0,\L,0) -- (0,\L,\L);

    % === Right Cube ===
    \begin{scope}[shift={(2,0,0)}]
        \draw[thick] (0,0,0) -- (\L,0,0) -- (\L,\L,0) -- (0,\L,0) -- cycle;
        \draw[thick] (0,0,\L) -- (\L,0,\L) -- (\L,\L,\L) -- (0,\L,\L) -- cycle;
        \draw[thick] (0,0,0) -- (0,0,\L);
        \draw[thick] (\L,0,0) -- (\L,0,\L);
        \draw[thick] (\L,\L,0) -- (\L,\L,\L);
        \draw[thick] (0,\L,0) -- (0,\L,\L);
    \end{scope}

    % === Interface Plane ===
    \fill[blue!20,opacity=0.5] (\L,0,0) -- (\L,\L,0) -- (\L,\L,\L) -- (\L,0,\L) -- cycle;

    % === 1D Line inside the cubes (across both cubes)
    % From x = 0 to x = 4 (2 cubes wide), centered in y and z
    \draw[very thick, black] (0,\L/2,\L/2) -- (4,\L/2,\L/2);

    % === Reduction Arrow
    \draw[->,thick] (5.1,\L/2,\L/2) -- ++(1.5,0,0);

    % === Coordinate Axes ===
    \draw[->, thick] (-1.5,-0.4,\L/2) -- ++(0.8,0,0) node[anchor=north east] {$x$};
    \draw[->, thick] (-1.5,-0.4,\L/2) -- ++(0,0.8,0) node[anchor=north west] {$y$};
    \draw[->, thick] (-1.5,-0.4,\L/2) -- ++(0,0,0.8) node[anchor=south] {$z$};

    % === 1D Reduced Line to the Right, same length, with center separator
    \begin{scope}[shift={(7.2,0,0)}]
        \draw[very thick, black] (0,\L/2,\L/2) -- (4,\L/2,\L/2);
        % Tick at the midpoint (interface between two cubes, x = 2)
        \draw[thick] (2,\L/2 - 0.2,\L/2) -- (2,\L/2 + 0.2,\L/2);
    \end{scope}
    
    \end{tikzpicture}
    \caption{Penalty parameter $\sigma$ in hexahedral elements using tensor product basis functions can be reduced to the one-dimensional case. The problem is projected onto the dimension normal to the shared interface between two elements, in this case the $x$-direction. The penalty from the one-dimensional reduction can be used directly for the original three-dimensional case.}
    \label{fig:pen3d}
\end{figure}

\subsection{Continuous projection} \label{sect:cont_proj}
Consider isolating the penalty matrix from the discrete Laplacian: 
\begin{equation}
    P_{ij} = \int_\Gamma [[\phi_i]] \cdot [[\phi_j]] ~ds,
\end{equation}
and note that the basis coefficients of any continuous function in $V_\mathrm{DG}$ lie in the null space of $P$. As such, given any function $v \in V_\mathrm{DG}$, we can obtain the continuous part via orthogonal projection onto the null space of $P$.

As $P$ is a symmetric positive semi-definite matrix, the projection $\Pi$ onto the null space of $P$ can be computed as 
\begin{equation}
    \Pi =  I - P^+ P = I - \lim_{\varepsilon \rightarrow 0} (P + \varepsilon I)^{-1} P =  \lim_{\varepsilon \rightarrow 0} \varepsilon (P + \varepsilon I)^{-1}. 
\end{equation}
% continuous projection can be computed as
% \begin{equation}
%     \Pi_P (v) = (I - P P^+)v,
% \end{equation}
% where $P^+$ denotes the Moore-Penrose pseudoinverse. To compute the action of the pseudoinverse, we can note that
% \begin{equation}
%     P^+ v = \lim_{\varepsilon \rightarrow 0} ( P + \varepsilon I)^{-1} v.
% \end{equation}
In practice, we pick a small value of $\varepsilon = 10^{-3}$ and evaluate the action of the limiting expression 
\begin{equation}
\Pi_\varepsilon := \varepsilon (P + \varepsilon I)^{-1}
\end{equation}
with a linear solve. To perform the linear solve, we apply the conjugate gradient algorithm preconditioned with an incomplete Cholesky factorisation with zero additional fill. We find that this  simple approach is sufficient for our numerical tests as it is only applied as a post-processing step and does not present as a computational bottleneck. Moreover, by contrast, incomplete Cholesky for our Laplacian solves performs poorly.

To understand how $\varepsilon$ controls the failure of continuity, let $c$ denote the vector of coefficients of an $L^2$-normalized function in our DG basis, so that $\Vert c \Vert_2 = 1$, and let $c_\varepsilon := \Pi_\varepsilon c$ denote the coefficients after projection. Then the `discontinuity seminorm' after projection can be bounded as 
\begin{equation}
\sqrt{ c_\varepsilon^\top P c_\varepsilon } \leq \varepsilon / 2.
\end{equation}
Indeed, $ c_\varepsilon^\top P c_\varepsilon = c^\top f_{\varepsilon} (P) c$, where $f_\varepsilon (P)$ denotes the matrix function induced by the univariate function $f_\varepsilon (x) = \varepsilon^2 x / (x + \varepsilon^2)$, which satisfies $f_\varepsilon (x) \leq \varepsilon / 4$ for all $x \geq 0 $. The discontinuity seminorm after projection may in fact be significantly smaller, since $c$ typically represents a function that is already nearly continuous.

In practice, we find that post-processing our solutions using such an approximate projection only negligibly affects our estimates of Hartree-Fock and DFT energies, but we still include the step due to the relatively insignificant cost and the reassurance of approximately restoring variational energy guarantees that are automatic for continuous Galerkin approaches.

Indeed, for the purpose of estimating energies, note that the action of the discrete DG Laplacian matrix coincides with the that of the continuous part 
\begin{equation}
    T_{ij}^\mathrm{cont} = \sum_{m=1}^M \int_{K_m} \nabla \phi_i(\mathbf{r}) \cdot \phi_j(\mathbf{r}) ~d\mathbf{r}.
\end{equation}
on the null space of $\Pi$.
After the approximate projection is applied to define new normalized basis coefficients ${c}_\varepsilon := \Pi_\varepsilon c $, we compute the kinetic energy using $T^{\mathrm{cont}}$ in place of $T$. In our implementation, we verify that
\begin{equation}
    \frac{ {c}_\varepsilon^\top (T-T^\mathrm{cont}) {c}_\varepsilon} { \Vert {c}_\varepsilon\Vert^2 } < 10^{-8}
\end{equation}
to ensure that the use of $T^\mathrm{cont}$ does not affect the calculation of energies.

\section{Discontinuous basis sets} \label{sect:bases}

We describe the choice of computational basis functions $\{ \phi_i, \ i=1,\ldots,N_\phi \}$ used to discretise the orbitals. As noted previously, all basis functions that we consider are linear combinations of tensor product functions that can be written 
$$f(x,y,z) = f(x)f(y)f(z).$$

To build the computational basis, we first construct a provisional discontinuous basis set $\{ \eta_i, \ i=1,\ldots,N_\eta \}$ by selecting on each element, functions from a predefined set of primitives. In our framework, primitive functions include polynomials and Gaussians, both of which satisfy the tensor product structure. To obtain the computational basis set, linear combinations of the primitives in the provisional set are chosen via an adaptive filtration procedure, resulting in a compact orthonormal basis set.

\subsection{Primitive functions}\label{sec:componentfunctions}

\subsubsection{Polynomials}
Perhaps the most common basis functions used in traditional DG methods are tensor product polynomials of the form
\begin{equation}
    f(x,y,z) = x^{p_x} y^{p_y} z^{p_z}, \ p_x,p_y,p_z \leq p,
\end{equation}
where the maximal polynomial degree $p \geq 0$ can be  chosen freely. Each element $K_m$ is typically equipped with its set of tensor product polynomials. Polynomials provide a systematic means of constructing a complete approximation space and are widely used in DG applications.

Although previous work has explored the use of polynomial bases in quantum chemistry \cite{kanungo2017large, rufus2021fast}, pure polynomial basis sets are less favoured in this field. The major reason is the presence of atomic cusps. Accurately resolving the sharp features of wavefunctions near atomic centres typically requires a very large number of polynomial basis functions, compared to more commonly used Gaussian-type orbitals.

When employed, polynomial basis functions are often represented either as orthogonal Legendre polynomials or via Lagrange interpolating polynomials using Gauss–Lobatto points. These representations improve the numerical conditioning of the resulting DG system matrices relative to the standard monomial basis. However, it is important to note that all these choices span the same underlying polynomial space.

Since the meshes considered in this work may extend to infinite domains, the support of polynomial basis functions on boundary elements of $\mathcal{T}$ must be truncated to ensure that they remain square-integrable. In our implementation, we restrict the support of any polynomial basis function to the bounding box $[a_x,b_x] \times [a_y,b_y] \times  [a_z,b_z]$ defined via \eqref{eq:bounding_box} in our mesh generation procedure.

\subsubsection{Gaussians} \label{sect:gaussian_basis}
One of the most popular type of basis functions in quantum chemistry calculations is the Gaussian-type orbital (GTO), defined as
\begin{equation}
    g(\mathbf{r}) = x^{l_x} y^{l_y} z^{l_z} \exp(-\alpha|\mathbf{r}-\mathbf{R}|_2^2), \quad l_x + l_y + l_z \leq l,
\end{equation}
where $\mathbf{r} = (x,y,z)$ is the spatial coordinate, $\mathbf{R}$ is the centre of the Gaussian, $\alpha > 0$ controls the width, and $l \geq 0$ is the total angular momentum quantum number.

%\ML{can write this better. Clarify you are describing a procedure to turn a single ordinary GTO into a collection of DG basis functions. Use letters, like explain how to define each $\phi$ as a GTO ($\psi$?) times an indicator} 

To furnish valid basis functions in a DG framework, we  consider the restrictions of a single GTO to several individual mesh elements near its centre, which altogether contain the GTO in their span. Specifically, for a given element $K_m$, we build an element-localised basis function $f(\mathbf{r})$ by multiplying a GTO $g(\mathbf{r})$ with an indicator function $\chi_{m}(\mathbf{r})$: 
\begin{equation}
    f(\mathbf{r}) = g(\mathbf{r}) \cdot \chi_{m}(\mathbf{r}),  \qquad  \chi_m (\mathbf{r}) = \begin{cases}
        1, & \mathbf{r} \in K_m, \\
        0, & \text{otherwise.} \\
    \end{cases}
\end{equation}
We construct such functions $f(\mathbf{r})$ only for elements $K_m$ that intersect the ball
\begin{equation}
    B_\alpha(\mathbf{R}) = \left\{ \mathbf{r} : | \mathbf{r} - \mathbf{R} | \leq {3} / {\sqrt{4\alpha}} \, \right\}.
\end{equation}
This procedure effectively decomposes a single GTO into a collection of DG basis functions with disjoint support, allowing the original function to be represented accurately across multiple elements while respecting the locality required by DG methods.

Unlike polynomial basis functions, GTOs are naturally square-integrable due to their exponential decay and require no additional truncation even when defined over unbounded domains.

\begin{figure}
    \centering
    \includegraphics[scale=0.35]{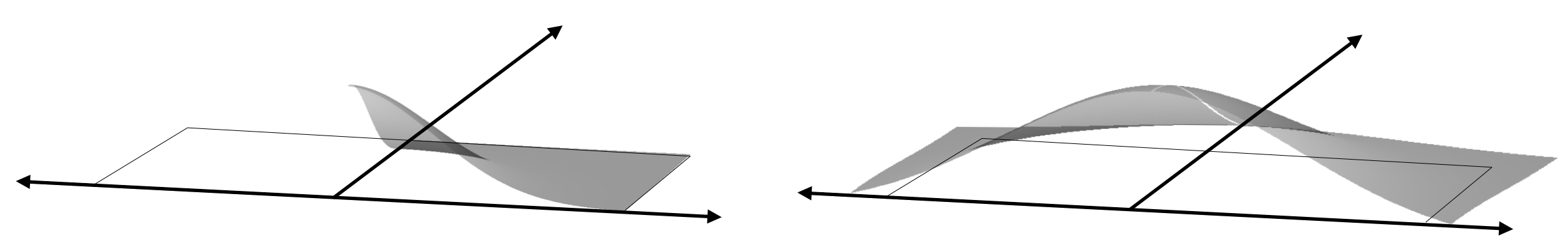}
    \caption{DG polynomial and Gaussian basis functions. Polynomial basis functions (left) on infinite domains are truncated to have support only within original mesh bounding box. Gaussian basis functions (right) are not truncated but are split across elements on which they have significant mass and duplicated.}
    \label{fig:basis_def}
\end{figure}

\subsection{Adaptive basis construction} \label{sect:adaptive_truncation}
To construct the DG basis, we begin by specifying, for each element $K_m$, a local basis set consisting of primitive functions following Section~\ref{sec:componentfunctions}. We refer to this as the provisional basis $\{ \eta_i, \ i = 1,\ldots,N_\eta \}$. A key advantage of the discontinuous Galerkin framework is its flexibility, allowing for arbitrary mixtures of basis functions to be chosen independently in each element for the provisional set.

We then filter the provisional basis. First, we orthogonalise the basis by performing a singular value decomposition (SVD) of the overlap matrix~\eqref{eq:overlap}, which is computationally efficient due to the block diagonal structure of the matrix. Within this orthogonalised provisional basis, we solve the one-electron eigenvalue problem
\begin{equation} \label{eq:oneH}
    \bigg( \Delta + \sum_I \frac{Z_I}{|\mathbf{r} - \mathbf{R}_I|} \bigg) \tilde{\psi}_k = \varepsilon_k \tilde{\psi}_k, \qquad k=0,\ldots, N_\mathrm{init}-1
\end{equation}
induced only by the kinetic energy and external potential. The number of states retained is defined following the discussion below in \eqref{eq:ninit}. Further details of the eigenfunction computation are deferred to Section \ref{sect:lobpcg}.

To construct the computational basis set $\{ \phi_i \}$ from the provisional set $\{ \eta_i \}$, we perform the following filtration procedure independently on each element $K_m, \ m=1,\ldots,M$:
\begin{enumerate}
    \item The global one-electron eigenfunctions $\tilde{\psi}_k$ are restricted to the domain of element $K_m$, yielding a set of localised functions $\tilde{\psi}_k|_{K_m}$.
    \item A local overlap matrix is constructed for each element $m$: 
    $$ [S_m]_{ij} = \int_{K_m} \tilde{\psi}_i|_{K_m} (\mathbf{r}) \tilde{\psi}_j|_{K_m} (\mathbf{r}) ~d(\mathbf{r}).$$
    \item An SVD of the local overlap matrix is performed: $S_m = U_m \Sigma_mV_m^T$.
    \item The top $N_\mathrm{filt}$ left singular vectors from $S_m$, corresponding to the largest singular values, are selected. These define orthonormal computational basis functions for element $K_m$.
\end{enumerate}
The number $N_\mathrm{filt}$ of retained functions  can be adjusted per element, enabling flexible, adaptive control over basis sizes. To ensure sufficient coverage of the global eigenfunction space, the number of eigenfunctions $N_\mathrm{init}$ retained in the precomputation is chosen as
\begin{equation} \label{eq:ninit}
    N_\mathrm{init} = C N_{\mathrm{filt}} N_e,
\end{equation}
where $C$ is a constant which we take to be $1$ for simplicity. Overall the cost of precomputation is outweighed by the SCF procedure downstream.

\section{Coulomb integrals} \label{sect:coulomb}
We now turn to the discretisation of operators \eqref{eq:HF} involving the Coulomb potential, namely the external potential and the two-electron potential. Although special care is needed to accurately compute the relevant integrals, the discontinuity of the basis set does not impose any fundamental concerns in the same way as it does for the discretisation of the Laplacian.

The tensor product structure of the primitive functions allows for fast computation of the Coulomb integrals. However, as the ultimate computational basis consists of linear combination of the primitives, this cannot be applied directly to compute the integrals. Instead we compute integrals in the provisional basis, and project into the computational basis to obtain the desired final values.

Analytical expressions are available for computing Coulomb integrals using GTO bases in free space calculations, and can be found for instance in \cite{szabo2012modern}. However, these formulas are not directly applicable to our basis sets, as element domains in the DG framework are strict subsets of $\mathbb{R}^3$. Furthermore, our strategy must be able to handle the case of polynomial basis functions as well as GTOs. We introduce a unified framework for the fast and highly accurate computation of the desired integral quantities. 

\subsection{Operator construction}

In this section we consider calculation of the matrix entries for the external potential \eqref{eq:TandU} and the two-body electron repulsion integrals \eqref{eq:Vijkl}. To compute these efficiently, we first form the corresponding integrals in our provisional basis: 
\begin{equation} \label{eq:Vijkl_tilde}
    \begin{split}
        \tilde{U}_{ij} = \sum_I \int_{\mathbb{R}^3} \eta_i(\mathbf{r}) &\frac{Z_I}{|\mathbf{r}-\mathbf{R}_I|} \eta_j(\mathbf{r}) ~d \mathbf{r}, \\
         \tilde{V}_{ijkl} = \int_{\mathbb{R}^3} \int_{\mathbb{R}^3} \eta_i (\mathbf{r}_1) \eta_j (\mathbf{r}_1) &\frac{1}{|\mathbf{r}_1-\mathbf{r}_2|} \eta_k (\mathbf{r}_2) \eta_l (\mathbf{r}_2) ~d\mathbf{r}_1 d\mathbf{r}_2,
    \end{split}
\end{equation}
The advantage of first constructing the operators in the provisional basis is that each of the functions $\eta_i$ is a pure tensor product, which enables reduction of each of the above integrals into one-dimensional integrals following a procedure that we outline below.

The desired integrals in the final computational basis can then be recovered as follows. Via Section~\ref{sect:adaptive_truncation} we can express each computational basis functions as a linear combination of the provisional basis functions
\begin{equation}
    \phi_i = \sum_{j=1}^{N_\eta} C_{ij} \eta_j.
\end{equation}
Then in terms of the coefficients $C_{ij}$ we compute
\begin{equation}
    \begin{split}
        U_{ij} = \sum_{k=1}^{N_\eta} \sum_{l=1}^{N_\eta} C_{ki} &\tilde{U}_{kl} C_{lj} \\
        V_{ijkl} = \sum_{k=1}^{N_\eta} \sum_{l=1}^{N_\eta} C_{ai} C_{bj} &\tilde{V}_{abcd} C_{ck} C_{dl}. \\
    \end{split}
\end{equation}
Therefore in the following we focus only on the computation of suitable integrals in the provisional basis.

\subsection{Gaussian sum approximation} \label{sect:gauss_approx}
We begin by addressing the evaluation of integrals involving the singular Coulomb kernel, which arise when constructing the external potential operator in Eq.~\eqref{eq:TandU}. Specifically, we consider integrals of the form
\begin{equation} \label{eq:I1}
    I_1 = \int_D \frac{ f(\mathbf{r}) }{\|\mathbf{r}\|}~d\mathbf{r},
\end{equation}
where $f(\mathbf{r}) = f_x(x)f_y(y)f_z(z)$ is a separable function defined on the domain $D$.

We adopt the strategy proposed in \cite{white2023nested}, which uses a sum of Gaussians to approximate the singular kernel
\begin{equation}
    \frac{1}{\|\mathbf{r}\|} \approx \sum_{k=1}^{N_G} c_k \exp(-\alpha_k \| \mathbf{r} \|^2 ).
\end{equation}
Details of this Gaussian sum approximation are provided in Appendix B of \cite{white2023nested}, where it is shown that on the order of $N_G = 50 \sim 100$ Gaussians suffice to achieve good accuracy. Alternative approaches to exponential sum approximation include \cite{beylkin2005approximation,beylkin2010approximation}.

Focusing on domains strictly of the form $D=[a_x,b_x] \times [a_y,b_y] \times [a_z,b_z],$ this approximation allows us to approximate the Coulomb integral \eqref{eq:I1} as 
\begin{equation}
    I_1 \approx \sum_{k=1}^{N_G} c_k \prod_{w \in \{x,y,z \} } \int_{a_w}^{b_w} \exp(-\alpha_k w^2 ) f_w(w) ~dw.
\end{equation}
This right-hand side can be evaluated efficiently as it reduces to the calculation of one-dimensional integrals.

This allows us to compute matrix entries $U_{ij}$ \eqref{eq:TandU} of the external potential in the provisional basis set by subsituting $f \rightarrow \eta_i \eta_j$ in to the above formula. Note that similar to the overlap matrix, the external potential exhibits block diagonal sparsity pattern where $U_{ij} \neq 0$ if and only if $\eta_i, \eta_j$ are supported on the same element.

This procedure reduces the evaluation of singular integrals to a sequence of one-dimensional Gaussian integrals. When basis functions $\eta_i,\eta_j$ are Gaussian, these integrals can be evaluated analytically using error functions. For polynomial basis functions, the computational details are provided in Appendix~\ref{appendix:gaussian}.

\subsection{Two-body integrals}

We can similarly use the Gaussian sum approximation to calculate two-body Coulomb integrals of the form 
\begin{equation}
    I_2 = \int_{D} \int_{D'}\ \frac{ f(\mathbf{r}) g(\mathbf{r'}) }{\|\mathbf{r} - \mathbf{r}'\|}~d\mathbf{r'} ~d\mathbf{r},
\end{equation}
where $f(\mathbf{r}) = f_x(x)f_y(y)f_z(z)$ and $g(\mathbf{r}) = g_x(x)g_y(y)g_z(z)$ are both separable. As before, we limit the domains of integration to be cubes, with $D$ defined as before and $D' = [a_x',b'_x] \times [a_y',b'_y] \times [a_z',b'_z]$. Evaluation of such integrals is required to calculate the electron repulsion integrals or ERIs \eqref{eq:Vijkl}. Although we do not directly use the ERIs in our implementation of HF and DFT, we note that it may remain important to form them for post-HF calculations.

As before, we apply the Gaussian sum approximation to decompose
\begin{equation} \label{eq:i2_gaussian}
    I_2 \approx \sum_{k=1}^{N_G} c_k \prod_{w \in \{x,y,z \} } \int_{a_w}^{b_w} \int_{a'_w}^{b'_w} \exp(-\alpha_k (w-w')^2 ) f_w(w) g_w(w') ~dw' ~dw.
\end{equation}
This however cannot be reduced directly to the calculation of one-dimensional integrals as the Gaussian term does not split.

To handle this difficulty we replace the Gaussians with Fourier cosine expansions 
\begin{equation} \label{eq:fourier_gaussian}
    g_\alpha(x) =\exp(-\alpha x^2 ) \approx \sum_{k=0}^{N_F} a_\alpha^{(k)} \cos\bigg(\frac{k\pi x}{L_\alpha}\bigg),
\end{equation}
where $L_\alpha$ is a constant to be set. Full details of the Fourier expansion are provided in Appendix \ref{appendix:fourier}. This additional expansion allows us to decompose the terms in \eqref{eq:i2_gaussian} using the cosine double angle formula as 
\begin{equation} \label{eq:i2_split}
    \begin{split}
        I_2 \approx \sum_{n=1}^{N_G} c_n \prod_{w \in \{x,y,z \} } \sum_{k=0}^{N_F} &a^{(k)}_{\alpha_n} \int_{a_w}^{b_w} \int_{a'_w}^{b'_w} \cos\bigg( \frac{k\pi (w-w')}{L_{\alpha_n}} \bigg) f_w(w) g_w(w') ~dw' ~dw \\
        = \sum_{n=1}^{N_G} c_{n} \prod_{w \in \{ x,y,z \} } \sum_{k=0}^{N_F} &a_{\alpha_n}^{(k)} \bigg( \int_{a_w}^{b_w} \cos\bigg(\frac{k\pi w}{L_{\alpha_n}} x\bigg) f_w(w) ~dw \cdot \int_{a'_w}^{b'_w} \cos\bigg(\frac{k\pi w'}{L_{\alpha_n}}\bigg) g_w(w') ~dw' \\
        &+ \int_{a_w}^{b_w} \sin\bigg(\frac{k\pi w}{L_{\alpha_n}}\bigg) f_w(w) ~dw \cdot \int_{a'_w}^{b'_w} \sin\bigg(\frac{k\pi w'}{L_{\alpha_n}}\bigg) g_w(w') ~dw' \bigg).
    \end{split}
\end{equation}
Entries of the electron repulsion integral in the provisional basis \eqref{eq:Vijkl_tilde} can be computed by substituting in $f \rightarrow \eta_i \eta_j$ and $g \rightarrow \eta_k \eta_l$ into the above formula. This expression involves only one-dimensional cosine product integrals which can be computed efficiently. Computational details of the last step are provided in Appendix \ref{appendix:cosine}. 

\section{Auxiliary grid evaluations} \label{sect:auxiliary_mesh}

While \eqref{eq:i2_split} enables direct computation of the electron repulsion integral $V_{ijkl}$ \eqref{eq:Vijkl}, the cost of forming the full four-index tensor (with nonzero entries scaling quadratically in the number of elements) can be avoided for mean-field methods, i.e., HF and DFT.

In the HF case, we need only the electron density $\rho$, the Hartree potential $V_\mathrm{H}$, and the Fock exchange $V_\mathrm{X}$. Fortunately, the matrices of $V_\mathrm{H}$ and $V_\mathrm{X}$ within our computational basis set $\{ \phi_j \}_{j=1}^{N_\phi}$ need not be constructed explicitly. Indeed, note that the matrix of the Fock exchange operator is dense and full-rank and as such can be prohibitively expensive to form and store.
Instead, we need only compute the action of these operators on arbitrary orbitals $u(\mathbf{r}) = \sum_{j=1}^{N_\phi} u_j \phi_j(\mathbf{r})$ in our computational subspace.

The action of these operators on a given orbital can be efficiently evaluated by solving several Poisson equations. Specifically, the Hartree potential requires a single Poisson solve, while the Fock exchange involves solving $N_e$ Poisson problems. Details of these reductions are provided in Section~\ref{sect:twoE}.

For DFT, we likewise need the electron density $\rho$ and the Hartree potential $V_\mathrm{H}$. Under the local density approximation (LDA), we additionally compute the exchange-correlation $V_\mathrm{XC}$, while the Fock exchange is omitted. 

The Poisson problems are not solved directly in the DG basis. Instead, they are discretised and evaluated using an auxiliary mesh, denoted $\mathcal{S}$, on which a real space interpolating grid is defined. The electron density and exchange-correlation are similarly represented pointwise on this grid. The use of this auxiliary representation is motivated by the fact that the DG basis used for molecular orbitals is generally not well suited for representing either the electron density or the solutions to the Poisson problems required for $V_\mathrm{H}$ and $V_\mathrm{X}$.

\subsection{Poisson equations for Hartree potential and Fock exchange} \label{sect:twoE}

We derive the Poisson equations that arise in the computation of the Hartree potential and Fock exchange operator. These equations are solved on an auxiliary mesh $\mathcal{S}$, described in more detail in Section \ref{sect:auxiliary_meshing}. The potentials are represented using interpolating grids defined on the auxiliary mesh, which also defines a Gauss quadrature rule for computing integrals relating to these quantities.

\subsubsection{Hartree potential}

The matrix entries $J_{ij}$ of the Hartree potential in \eqref{eq:HF} are given by
\begin{equation}\label{eq:hartreecoulomb}
    \begin{split}
        J_{ij} &= \int_{\mathbb{R}^3} \phi_i(\mathbf{r}) \phi_j(\mathbf{r}) \cdot \int_{\mathbb{R}^3}\frac{\rho(\mathbf{r}')}{|\mathbf{r}-\mathbf{r}'|} ~d\mathbf{r}' ~d\mathbf{r} \\
        &= \int_{\mathbb{R}^3} \phi_i(\mathbf{r}) \phi_j(\mathbf{r}) \cdot v_\mathrm{H}(\mathbf{r}) ~d\mathbf{r}
    \end{split}
\end{equation}
where the Hartree potential $v_\mathrm{H}$ is obtained by solving the Poisson equation
\begin{equation} \label{eq:hartree_poisson}
    -\Delta v_\mathrm{H}(\mathbf{r}) = 4\pi \rho(\mathbf{r}),
\end{equation}
on the full domain $\mathbb{R}^3$ with zero boundary condition at infinity. The Poisson solve is performed on the auxiliary mesh $\mathcal{S}$ following a procedure to be described below in Section \ref{sect:auxiliary_poisson}, yielding a suitable representation of $v_\mathrm{H}$ on the interpolating grid.

Rather than assembling the matrix $J_{ij}$ explicitly, we compute the action of the Hartree operator on an arbitrary orbital $u(\mathbf{r}) = \sum_{j=1}^{N_\phi} u_j \phi_j(\mathbf{r})$ as
\begin{equation}
        \sum_{j=1}^{N_\phi} J_{ij}u_j  = \int_{\mathbb{R}^3} \phi_i(\mathbf{r}) \cdot u(\mathbf{r}) v_\mathrm{H}(\mathbf{r}) ~d\mathbf{r}.
\end{equation}
This integral is evaluated numerically using the Gauss quadrature rule defined by the auxiliary grid. Each factor in the integrand is evaluated pointwise at quadrature points.

\subsubsection{Fock exchange}
The entries of the non-local Fock exchange matrix in \eqref{eq:HF} are 
\begin{equation} \label{eq:fock_exchange}
    K_{ij} = \int_{\mathbb{R}^3} \phi_i(\mathbf{r}) \cdot \sum_{s=1}^{N_e} \psi_s(\mathbf{r}) \int_{\mathbb{R}^3} \frac{\phi_j(\mathbf{r}') \psi_s(\mathbf{r}')}{|\mathbf{r} - \mathbf{r}'|} ~d\mathbf{r}' ~d\mathbf{r}.
\end{equation}
As with the Hartree term, we do not assemble the exchange matrix explicitly. Instead, we compute its action on an orbital $u(\mathbf{r}) = \sum_{j=1}^{N_\phi} u_j \phi_j(\mathbf{r})$, which yields
\begin{equation}
    \begin{split}
        \sum_j K_{ij}u_{j} = \int_{\mathbb{R}^3} \phi_i(\mathbf{r}) &\cdot  \sum_{s=1}^{N_e} \psi_s (\mathbf{r}) \int_{\mathbb{R}^3} \frac{u (\mathbf{r}') \psi_s (\mathbf{r}')}{|\mathbf{r}-\mathbf{r}'|} ~d\mathbf{r}' ~d\mathbf{r} \\
        = \int_{\mathbb{R}^3} \phi_i(\mathbf{r}) &\cdot  \sum_{s=1}^{N_e} \psi_s(\mathbf{r})v_{s} (\mathbf{r}) ~d\mathbf{r}
    \end{split}
\end{equation}
where each $v_{s}(\mathbf{r})$ solves the Poisson equation
\begin{equation} \label{eq:fock_poisson}
    -\Delta v_{s}(\mathbf{r}) = 4\pi u(\mathbf{r}) \psi_s(\mathbf{r}),
\end{equation}
on the full domain $\mathbb{R}^3$ with zero boundary condition at infinity. These $N_e$ Poisson problems are solved on the auxiliary mesh $\mathcal{S}$, following a procedure to be described below in Section \ref{sect:auxiliary_poisson}. As in the case of the Hartree potential, the integral is evaluated numerically via the Gauss quadrature rule on the auxiliary grid, with each factor in the integrand evaluated pointwise at quadrature points.

\subsection{Auxiliary mesh} \label{sect:auxiliary_meshing}

We outline the procedure to construct  the auxiliary mesh $\mathcal{S} = \{ J_m, \ m=1,\ldots,M_\mathrm{aux} \}$. In contrast with the orbital mesh $\mathcal{T}$, the auxiliary mesh is built using adaptive refinement to generate a spatial interpolating grid that permits pointwise operations.

The construction begins by generating a new mesh following the same procedure used for orbital mesh in Section \ref{sect:meshing}, but with a different bounding domain that captures the support of the basis functions in $\mathcal{T}$. Specifically, for each coordinate direction, we define the bounding interval as the union of all regions where the corresponding one-dimensional basis functions of the provisional basis are numerically significant:
\begin{equation} \label{eq:meshbound}
    [\tilde{a}_x, \tilde{b}_x] = \bigcup_i ~\{ x : |\eta_{i}^x(x)| > 10^{-9} \}
\end{equation}
and similarly for the $y$- and $z$-directions, where $\eta_{i}^x$ denotes the $x$-component of the tensor product provisional basis function $\eta_i$.
This choice ensures that the mesh spans only the region where the basis functions contribute meaningfully, avoiding unnecessary extension beyond their support.

\begin{figure}
    \centering
    \includegraphics[scale=0.4]{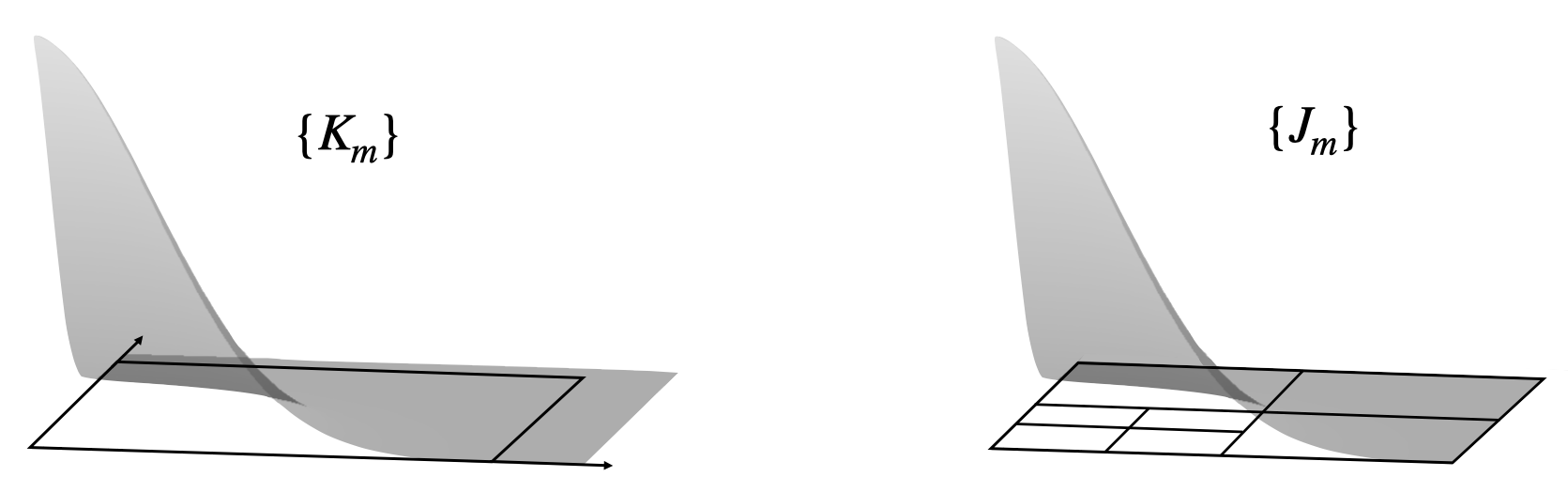}
    \caption{Schematic of adaptive mesh construction for solving Hartree and Fock exchange Poisson problems in two dimensions. The auxiliary mesh $\mathcal{S}$ with elements $\{ J_m \}$ (right) is constructed from the original DG mesh for the orbitals $\mathcal{T}$ with elements $\{ K_m \}$ (left) by truncating the domain to a rectangular region where $|\eta_i| > 10^{-9}$. The new auxiliary mesh $\mathcal{S}$ is then adaptively refined around sharp regions of the densities.}
    \label{fig:adaptive_mesh}
\end{figure}

The auxiliary mesh $\mathcal{S}$ is then adaptively refined using an octree-based structure. Initially, each element $J_m$ is equipped with piecewise degree-$q$ nodal tensor product polynomials. In practice, we find that a moderate polynomial degree $q=4$ offers a good balance between accuracy and computational cost.

The refinement is guided by the electron density $\rho(\mathbf{r})$. This is as the density is represented pointwise using an interpolating grid on $\mathcal{S}$, and is required in the Poisson solve for the Hartree potential \eqref{eq:hartree_poisson}. On each element $J_m$, we evaluate $\rho(\mathbf{r})$ at the Gauss–Legendre points $\{ \mathbf{x}_{im} \}_{i=1}^{(q+1)^3}$ and construct a local interpolant
\begin{equation} \label{eq:density_approx}
    \sum_{i=1}^{(q+1)^3} \rho(\mathbf{x}_{im}) P_i(\mathbf{r}),
\end{equation}
where $P_i(\mathbf{r})$ are the Lagrange interpolating polynomials associated with the Gauss–Legendre nodes.

To determine whether an element $J_m$ requires further refinement, we subdivide it uniformly into $n_\mathrm{ref}$ child elements denoted $\{H_j\}_{j=1}^{n_\mathrm{ref}}$, each similarly equipped with Gauss-Legendre points $\{ \tilde{\mathbf{x}}_{ij} \}_{i=1}^{(q+1)^3}$. On each child element $H_j$, we repeat the interpolation process using newly sampled values $\rho(\tilde{\mathbf{x}}_{ij})$.

\begin{figure}
    \centering
    \includegraphics[scale=0.4]{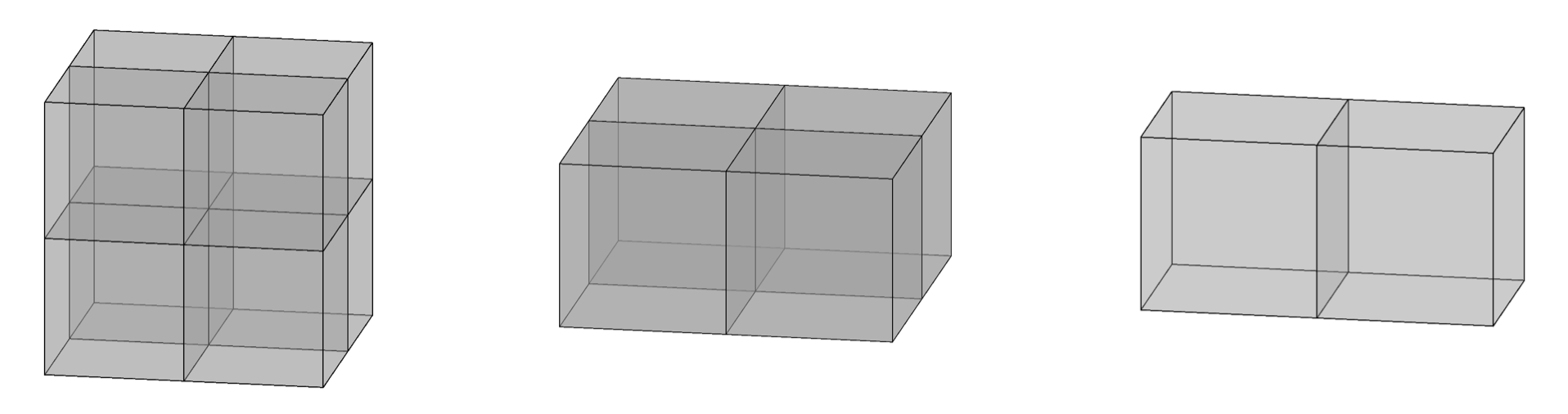}
    \caption{Visualisation of element refinement for auxiliary mesh $\mathcal{S}$. Elements with roughly equal side lengths are refined uniformly into 8 smaller boxes, otherwise elements are not divided in dimensions where the edge length is below half that of those in the other dimensions.}
    \label{fig:adaptive_ref}
\end{figure}

The number of subdivisions $n_\mathrm{ref}$
depends on the element’s aspect ratio. Letting $L$ denote the maximum of the edge lengths $L_x, L_y, L_z$ along each dimension of the element $J_m$, no subdivision occurs along the $x$-direction if $L_x < \frac{1}{2}L$, and likewise for $y,z$. Consequently, elements can be subdivided into $n_\mathrm{ref}=$ 2, 4, or 8 subelements depending on which spatial directions are refined, as illustrated in Figure \ref{fig:adaptive_ref}.

An error metric $Q_m$ is computed by comparing the integral of the interpolant on the original element with the sum over its children
\begin{equation}
    Q_m = \bigg| \sum_{i=1}^{(q+1)^3} \rho(\mathbf{x}_{im}) w_i - \sum_{j=1}^{n_\mathrm{ref}} \sum_{i=1}^{(q+1)^3} \rho(\tilde{\mathbf{x}}_{ij}) \tilde{w}_i \bigg|,
\end{equation}
where $w_i,\tilde{w}_i$ are the Gauss-Legendre quadrature weights on $J_m$ and its children respectively. This quantity estimates how accurately the density integral is captured on $J_m$. Although it does not account for integration against test functions directly, we find empirically that it yields results nearly identical to more sophisticated estimators based on test function projections. 

If the error indicator for an element $Q_m > \delta_q$, where $\delta_q$ is a user-specified tolerance, the element $J_m$ is replaced in $\mathcal{S}$ by its children, and the procedure repeated recursively. This refinement procedure is applied until all elements $J_m$ satisfy the criterion $Q_m < \delta_q$. In our implementation, we set a stringent tolerance of $\delta_q = 10^{-8}$. The final grid on $\mathcal{S}$ is then given by the collection of Gauss–Legendre points across all elements $J_m$, which are also used to define numerical integration rules on this auxiliary mesh.

\subsection{Poisson discretisation} \label{sect:auxiliary_poisson}
We discretise the Poisson problem using the same degree-$q$ nodal polynomial basis functions $P_i(\mathbf{r})$ in \eqref{eq:density_approx}. The Laplacian is constructed in the SIPDG framework to account for the possible discontinuities in the density $\rho(\mathbf{r})$ across the element boundaries of $\mathcal{T}$. Since the basis consists solely of piecewise polynomials, the penalty parameter for the DG Laplacian matrix can be chosen directly according to \eqref{eq:pen_poly}.

%\ML{I find this section a little sloppy. Logical ordering is also a bit confusing. it's not clear yet how it is motivated by coulomb integrals so the boundary conditions are confusing. Can you rewrite treating the boundary condition as given and pointing to future section where the boundary condition will be derived?} 

A remaining challenge is specifying appropriate boundary conditions for the Poisson problem. Indeed, the Poisson problems as stated in \eqref{eq:hartree_poisson} and \eqref{eq:fock_poisson} are formulated on all of $\mathbb{R}^3$, while the computational domain of the auxiliary mesh is finite. To handle this, we impose inhomogeneous Dirichlet boundary conditions on the boundary $\partial \Omega$ of the domain $\Omega := \bigcup_{m=1}^{M_{\mathrm{aux}}} J_m$ encompassed by the mesh $\mathcal{S} = \{ J_m : m = 1,\ldots ,M_{\mathrm{aux}} \}$.

Specifically, consider a general Poisson problem 
\begin{equation}
    -\Delta u = 4\pi f,
\end{equation}
over the infinite domain $\mathbb{R}^3$ with zero boundary condition at infinity and where $$f(\mathbf{r}) = \sum_{i=1}^{N_\phi} f_{ij} \phi_i(\mathbf{r})\phi_j(\mathbf{r})$$ is a function represented as a linear combination of pair products of the computational basis functions, as is the case for the right-hand sides of \eqref{eq:hartree_poisson} and \eqref{eq:fock_poisson}. Such a problem can be reformulated equivalently  on the finite domain  $\Omega $ as
\begin{equation}
    -\Delta u = 4\pi f \ \ \mathrm{in} \  \Omega \, ;  \quad ~u(\mathbf{r}) = \int_{\Omega } \frac{f(\mathbf{r}')}{\|\mathbf{r} - \mathbf{r}'\|} \, d\mathbf{r}', \ \ \mathbf{r} \in \partial \Omega.
\end{equation}
To compute the boundary condition, values of $f$ are interpolated onto the grid on $\mathcal{S}$, and the integral is evaluated using the associated Gauss quadrature rule. This approach is accurate because the source function $f(\mathbf{r})$ vanishes up to machine precision on the boundary $\partial \Omega$ by construction of the auxiliary mesh \eqref{eq:meshbound}. As a result, the integrand remains smooth numerically for $\mathbf{r}$ near the boundary.

\subsection{Linear solver}
To solve the resulting DG Poisson problem, we employ the Conjugate Gradient algorithm preconditioned with standard hp-multigrid techniques \cite{antonietti2015multigrid,nastase2006high} on the adaptive mesh. In our hp-multigrid implementation, we first coarsen the polynomial degree $q$ until $q=1$, before coarsening the mesh geometrically using the nested hierarchy generated during the adaptive mesh construction. 

\subsection{Exchange-correlation} \label{sect:xc_lda}

For the DFT calculations, the exchange–correlation potential is evaluated on the same auxiliary grid. Under the local density approximation (LDA), the exchange-correlation potential is treated as a local functional of the electron density. In this work, we employ the LDA \cite{bloch1929bemerkung,dirac1930note} with the Vosko-Wilk-Nusair (VWN) correlation energy \cite{vosko1980accurate}, as implemented by default in PySCF.
More advanced functionals could be treated analogously, but our focus here remains on this simple and widely used choice.

\section{Algorithmic pipeline} \label{sect:alg_overview}

\subsection{Eigenvalue solver} \label{sect:lobpcg}
Each self-consistent field (SCF) iteration requires the solution of an eigenvalue problem involving the Fock or Kohn–Sham operators. To solve these problems, we employ the Locally Optimal Block Preconditioned Conjugate Gradient (LOBPCG) algorithm \cite{knyazev2001toward}.

We choose LOBPCG because it supports preconditioning of the eigenvalue problem, which is essential for ensuring that the number of iterations remains stable as the basis is refined and the condition number of the discretised Laplacian increases. In our implementation, we use adaptive smoothed aggregation ($\alpha$SA) multigrid as a preconditioner, tailored for DG discretisations \cite{pan2025geometric}. The preconditioner is constructed for the operator
\begin{equation}
T + U - 2\tilde{\varepsilon}_0 I,
\end{equation}
where the shift $\tilde{\varepsilon}_0$ corresponds to the lowest eigenvalue from the one-electron problem \eqref{eq:oneH}. This shift ensures that the preconditioner is built for a symmetric positive definite operator. As the preconditioner remains fixed throughout the SCF procedure, it need only be constructed once during an initial offline setup phase.

Importantly, a single application of the adaptive multigrid preconditioner achieves $O(MN_\mathrm{filt}^2)$ cost, independently of the choice of basis functions. Standard hp-multigrid techniques do not directly apply to basis sets such as ours which consists possibly of a mixture of polynomials and Gaussian-type primitives on each element. In Section \ref{sect:ex_precond}, we provide an experimental evaluation of the performance of the preconditioned eigensolver.

\subsection{Self-consistent field iteration} \label{sect:scf}

We describe the SCF procedures used in our implementation for solving the HF \eqref{eq:HF} and Kohn-Sham DFT \eqref{eq:KS} equations. 

For HF, we use the Adaptively Compressed Exchange (ACE) technique \cite{lin2016adaptively,lin2019convergence} for speeding up the iterations. Instead of forming the full exchange matrix $K$ from \eqref{eq:fock_exchange}, the ACE method builds a rank-$N_e$ adaptively compressed Fock exchange operator $V^\mathrm{ACE}_{\mathrm{X}}$ which exactly matches the action of $V_\mathrm{X}$ on the occupied subspace.
% satisfying
% \begin{equation} \label{eq:Ktilde}
%    (V^\mathrm{ACE}_X \psi_k)(\mathbf{r}) = (V_X \psi_k)(\mathbf{r}), \quad k = 1,\ldots,N_e
% \end{equation}
% where $\psi_k$ are the occupied molecular orbitals.
For specifics of this construction we refer the reader to the original publication. The adaptively compressed operator is only updated periodically in the SCF procedure and reused until the electron density converges within a numerical tolerance, at which point the ACE representation is rebuilt.

The overall ACE procedure adopts a two-loop structure: an outer loop that rebuilds the ACE representation as needed and an inner loop that iterates the electron density and the orbitals to self-consistency for a fixed adaptively compressed exchange operator. The pseudocode below summarises this process. 

\begin{algorithm}[H]
\caption{Two-Loop Hartree--Fock SCF with ACE}
\label{alg:hf-scf-ace}
\begin{algorithmic}[1]
\While {exchange not converged}
    \State Construct adaptively compressed exchange $V^\mathrm{ACE}_{\mathrm{X}}$
        \While {electron density not converged}
        \State Solve eigenvalue problem \eqref{eq:HF}, with $V^\mathrm{ACE}_{\mathrm{X}}$ in place of $V_{\mathrm{X}}$, for MOs $\psi_i,\, i = 1, \ldots, N_e$,
        \State Update the density $\rho$ from the molecular orbitals
    \EndWhile
\EndWhile
\end{algorithmic}
\end{algorithm}

For Kohn-Sham DFT using the LDA exchange-correlation functional, the SCF procedure involves only the Hartree and local exchange-correlation potentials and thus requires no ACE construction or two-loop structure. The single-loop algorithm is summarised below.
\begin{algorithm}[H]
\caption{Kohn-Sham DFT SCF with LDA approximation}
\label{alg:lda-scf}
\begin{algorithmic}[1]
\While {electron density not converged}
    \State Solve eigenvalue problem \eqref{eq:KS} for KS orbitals $\psi_i,\, i = 1, \ldots, N_e$
    \State Update the density $\rho$ from the KS orbitals
\EndWhile
\end{algorithmic}
\end{algorithm}

In both SCF procedures, convergence of the Hartree potential is accelerated using the Anderson acceleration method \cite{anderson1965iterative}. This method is also widely known in electronic-structure theory as Pulay mixing or direct inversion in the iterative subspace (DIIS) \cite{pulay1980convergence}.

\subsection{Overview}
Here we outline all the steps of the entire algorithmic framework and their computational complexities. The symbols used in the overview are summarised in Table \ref{tab:constants} for reference.

\begin{table}[h]
    \centering
    \begin{tabular}{ll}
    \toprule
    \textbf{Symbol} & \textbf{Description} \\
    \midrule
    $N_e$           & Number of electrons \\
    \midrule
    \multicolumn{2}{c}{\textit{Molecular orbital mesh constructs}} \\
    \midrule
    $\mathcal{T}$   & Molecular orbital (MO) mesh \\
    $K_m$   & Element on MO mesh \\
    $M$     & Number of elements \\
    $N_\phi$        & Total number of basis functions\\
    $N_\mathrm{prim}$   & Number of primitive basis functions per element \\
    $N_\mathrm{trunc}$  & Number of basis functions per element post truncation \\
    \midrule
    \multicolumn{2}{c}{\textit{Auxiliary mesh constructs}} \\
    \midrule
    $\mathcal{S}$ & Auxiliary mesh \\
    $M_{\mathrm{aux}}$     & Number of elements in auxiliary mesh $\mathcal{S}$ \\
    $q, \, Q := q^3$             & Polynomial degree used on auxiliary grid \\
    \bottomrule
    \end{tabular}
    \caption{Definition of symbols used in algorithm overview.}
    \label{tab:constants}
\end{table}

\begin{enumerate}
    \item The user specifies the atomic positions $\{ \mathbf{R}_I \}$. A mesh $\mathcal{T}$ is generated to discretise the orbitals $\psi_j$ according to the procedure in Section \ref{sect:meshing}.

    \textbf{Complexity:} Mesh generation scales linearly with the total number of atomic nuclei. Denoting the number of elements in $\mathcal{T}$ as $M$, this step scales as $O(M)$.
    
    \item For each element $K_m$, we specify a selection of the primitive basis functions as outlined in Section \ref{sec:componentfunctions}. In particular we can consider arbitrary collections of polynomials and GTOs, each with support restricted to an individual element. An effective orthonormal basis is obtained for each element by taking an SVD of the overlap matrix \eqref{eq:overlap}.

    \textbf{Complexity:} We denote the number of primitive basis functions on each element as $N_\mathrm{prim}$. This step scales as $O(MN_\mathrm{prim}^3)$ due to the element-wise SVD.

    \item DG matrices corresponding to the one-electron operators are assembled, and the one-electron eigenproblem
    \begin{equation}
        (T + U)\tilde{\psi}_k = \varepsilon_k \tilde{\psi}_k, \ k=1,\ldots,N_\mathrm{init}
    \end{equation}
    is solved. We use these one-electron eigenvectors as initial guesses for the SCF iterative procedure, as well as for the next adaptive truncation step.

    \textbf{Complexity:} The eigenvalue problem is of size $MN_\mathrm{prim}$. Assuming the adaptive multigrid preconditioned eigensolver converges in a constant number of iterations, the cost scales as $O(MN_\mathrm{prim}^2 N_\mathrm{init} + MN_\mathrm{prim} N_\mathrm{init}^2 )$.
    
    \item On each mesh element $K_m$ of the orbital mesh $\mathcal{T}$, the adaptive filtration procedure in Section \ref{sect:adaptive_truncation} is applied using the one-electron eigenfunctions $\tilde{\psi}_k$ to generate the orthonormal single-particle basis set $\{ \phi_i : \ i=1,\ldots,N_\phi \}.$ 

    \textbf{Complexity:} This step requires a local SVD on each element. The complexity of this operation in $O(MN_\mathrm{init}^3)$.
    
    \item The electron density $\rho(\mathbf{r})$ is constructed from the one-electron eigenfunctions $\tilde{\psi}_k$. Following Section \ref{sect:auxiliary_mesh}, the auxiliary mesh $\mathcal{S}$ is constructed from this density, consisting of $M_{\mathrm{aux}}$ elements. This auxiliary mesh is reused throughout the SCF algorithm.

    \textbf{Complexity:} The exact value of $M_{\mathrm{aux}}$ is difficult to bound \emph{a priori} as it depends on the geometry of the molecule. For instance, the auxiliary mesh for a long chain molecule compared to for a compact geometry will likely require more elements, even for a fixed number of electrons $N_e$.

    To construct the auxiliary mesh, the electron density needs to be evaluated at each of the $O(M_{\mathrm{aux}} Q)$ auxiliary grid points. This incurs a cost of $O(M_{\mathrm{aux}} Q N_\mathrm{filt}),$ as every basis function $\phi_i$ within an element $K_m$ must be evaluated at each grid point.
    
    \item The HF equations are then solved self-consistently using the truncated basis until convergence as outlined in Section \ref{sect:scf}.

    \textbf{Complexity:} Assume that each LOBPCG call converges in a constant number of iterations and moreover that all SCF loops converge in a constant number of iterations. Moreover assume that the hp-multigrid preconditioned solver for each Poisson solve converges in a constant number of iterations. Then the overall complexity of the SCF procedure for is $O(M N_{\mathrm{filt}}^2 N_e + M N_{\mathrm{filt}} N_e^2 + M_{\mathrm{aux}} Q N_e^2)$ for HF and 
    $O(M N_{\mathrm{filt}}^2 N_e + M N_{\mathrm{filt}} N_e^2 + M_{\mathrm{aux}} Q )$ for DFT.

    % For methods requiring the Fock exchange, the cost per iterations then scales as $O(Mq^3N_e^2 )$, as $N_e^2$ Poisson solves are needed when applying the ACE strategy.

    % Otherwise, the Hartree potential term scales as $O(Mq^3)$. This is as only one Poisson solve is needed to resolve this. 

    % The final step of the SCF procedure requires solving a linear eigenvalue problem. Again, assuming the adaptive multigrid preconditioned eigensolver achieves linear scaling, this scales as $O(MN_\mathrm{filt}N_e^2),$ as in this case we want to solve for $N_e$ eigenvalues.
    
    % \item Finally, the HF energy and orbitals are returned.
\end{enumerate}

In future work considering post-HF methods, we anticipate that we can apply further adaptive truncation in terms of the converged Fock operator before pursuing downstream calculations. 

\section{Numerical results} \label{sect:results}

We present numerical results for several molecules. In all cases, the computational domain is treated as unbounded. Calculations are carried out using both Hartree–Fock (HF) and Density Functional Theory (DFT) with the local density approximation (LDA) functional.

To validate our approach, we benchmark our results against reference calculations performed with the PySCF package \cite{sun2018pyscf}. The numerical data for the Gaussian-type orbital basis sets used in these comparisons are obtained directly from the Basis Set Exchange \cite{pritchard2019new}. For the GTO bases considered, we use their uncontracted variants, meaning each Gaussian is treated as its own primitive function, in all of the numerical tests. All reference values where provided are computed using the highly accurate cc-pV5Z basis using PySCF.

\subsection{Adaptive multigrid preconditioning} \label{sect:ex_precond}
We first investigate the efficacy of the adaptive-multigrid-preconditioned LOBPCG solver deployed for the linear eigenvalue problems in the SCF procedure, cf. Section \ref{sect:lobpcg}. For this test we set up a linear chain of $n$ hydrogen atoms, spaced two Bohr radii apart. A DG mesh is defined with $n$ elements such that each element contains a single hydrogen atom.

A provisional basis is constructed using cc-pVTZ basis functions for each hydrogen atom, with supports suitably restricted to individual elements following Section \ref{sect:gaussian_basis}. The basis is then filtered following Section \ref{sect:adaptive_truncation} such that each element yields $N_\mathrm{filt}$ functions in the computational basis. We consider the values $N_\mathrm{filt} = 10, 20$ in our experiments.

Then we perform the first iteration of the SCF procedure (cf. Section \ref{sect:scf}) for solving the Kohn-Sham equations \eqref{eq:KS} and report the number of LOBPCG iterations required within this SCF iteration to reach a relative residual tolerance of $10^{-7}$. In Figure \ref{fig:adaptiveMG_hchain}, we report the number of LOBPCG iterations both using $\alpha$SA adaptive multigrid as a preconditioner and without any preconditioner at all. Using the $\alpha$SA preconditioner, we observe that the number of iterations increases only weakly with the length $n$ of the hydrogen. The weak dependence of the iteration count on system size suggests that the proposed solver scales well for larger systems. By contrast, the number of iterations scales quite poorly in the unpreconditioned case.

\begin{figure}[h]
    \centering
    \begin{tabular}{ |c||c|c|c|c|  }
    \hline
    \multicolumn{1}{|c||}{} & \multicolumn{2}{c|}{$N_\mathrm{filt}$ = 10} & \multicolumn{2}{c|}{$N_\mathrm{filt}$ = 20} \\
     \hline
     $n$ & $\alpha$SA & base & $\alpha$SA & base \\
     \hline
     2 & 9 & 67 & 10 & 131 \\
     4 & 14 & 278 & 14 & 374 \\
     8 & 17 & 1091 & 24 & 1528 \\
     16 & 23 & 1439 & 33 & 2194 \\
     \hline
    \end{tabular}
    \caption{Number of preconditioned LOBPCG iterations required to convergence within a single SCF iteration for the Kohn-Sham equations \eqref{eq:KS}. Here $n$ denotes the length of the hydrogen chain.}
    \label{fig:adaptiveMG_hchain}
\end{figure}

\subsection{Adaptive basis truncation} \label{sect:finding_ntrunc}

In each of the experiments below, we specify a series of primitive functions to serve as the provisional basis. These functions are then adaptively filtered following the procedure in Section \ref{sect:adaptive_truncation} to obtain the DG basis used in the calculations. Before presenting the results, we outline the procedure here used in the numerical experiments to determine the number of basis functions $N_\mathrm{filt}$ retained in each DG element following the truncation procedure.

For each of the molecules that we consider, we set up a experiment where a provisional basis is constructed using cc-pVQZ basis functions for each atom, with supports suitably restricted to individual elements.  Then we perform a HF or DFT calculation using this full DG basis with no filtration, furnishing a reference energy $E_0$. Finally, we apply the procedure in Section \ref{sect:cont_proj} to obtain an approximately continuous wavefunction which is used to evaluate the energy.

We note in all our numerical experiments that with our choice of penalty parameter following the discussion in Section \ref{sect:penalty}, application of the continuous projection does not alter the obtained wavefunctions by more than $10^{-5}$ in $L^2$ norm in practice. As such we only report the energies associated with the continuous projected wavefunctions.

To determine how many basis functions to retain after filtration, we run a series of simulations using the cc-pVQZ provisional basis, sweeping over $N_\mathrm{filt}$ to determine the smallest value such that the obtained energy differs from $E_0$ by less than $10^{-6}$. The smallest such number is reported and adopted as the value $N_\mathrm{filt}$ for adaptive filtration for experiments on that molecule. 

In the case where a provisional basis on a given element has fewer than $N_\mathrm{filt}$ functions, no filtration is performed on that element.
% This occurrence is due to lightweight bases being employed in some of the experiments.

\subsection{$\boldsymbol{\mathrm{H}_2}$}

We first consider the hydrogen molecule, $\mathrm{H}_2$. The two hydrogen atoms are positioned two Bohr radii apart, and the computational domain is divided into two elements, each containing one $\mathrm{H}$ atom. The setup is shown in Figure \ref{fig:hydrogen_setup}. Following the procedure described above in Section \ref{sect:finding_ntrunc}, we choose a maximum of $N_\mathrm{filt} = 50$ basis functions for each element. We consider two types of basis sets as the provisional basis: (1) GTOs only and (2) GTOs augmented with polynomial functions, for both HF and DFT calculations.  All supports are suitably restricted to individual elements, cf. Section \ref{sec:componentfunctions} for details.

\begin{figure}[h]
  \centering

  % Left: TikZ diagram
  \begin{minipage}[c]{0.5\textwidth}
    \centering
    \begin{tikzpicture}

      % Styles
      \tikzset{
        hydrogen/.style={circle, draw=black, fill=white, minimum size=8mm, inner sep=0pt},
        bond/.style={line width=1.2pt},
        box/.style={draw=blue, thick}
      }

      % Atom coordinates
      \coordinate (H1) at (-1, 0);
      \coordinate (H2) at ( 1, 0);

      % Bond
      \draw[bond] (H1) -- (H2);

      % Atoms
      \node[hydrogen] (H1_atom) at (H1) {\textbf{H}};
      \node[hydrogen] (H2_atom) at (H2) {\textbf{H}};

      % Bounding box dimensions
      \def\xmin{-2}
      \def\xmax{2}
      \def\ymin{-1}
      \def\ymax{1}
      \pgfmathsetmacro{\boxwidth}{(\xmax - \xmin)/2}

      % Two split rectangles
      \draw[box] (\xmin, \ymin) rectangle ++(\boxwidth, {\ymax - \ymin});
      \draw[box] (\xmin+\boxwidth, \ymin) rectangle ++(\boxwidth, {\ymax - \ymin});

    \end{tikzpicture}
  \end{minipage}%
  % Right: Table
  \begin{minipage}[c]{0.45\textwidth}
    \centering
    \begin{tabular}{@{}lll@{}}
      \toprule
      Atom & Coordinates \\
      \midrule
      H & (-1.0, 0.0, 0.0) \\
      H & (1.0, 0.0, 0.0) \\
      \bottomrule
    \end{tabular}
  \end{minipage}

  \caption{Mesh setup for $\text{H}_2$ molecule. Domain is split into two elements with atom positions in Bohr radii shown on the right.}
  \label{fig:hydrogen_setup}
\end{figure}

For the GTO-only case, the provisional basis equips each element with cc-pV$n$Z GTOs associated to both hydrogen atoms. We consider $n=2,3,4$, i.e., the cases of double (D), triple (T), and quadruple (Q) zeta functions. We compare the energies obtained with those from using the corresponding cc-pV$n$Z GTO basis in PySCF. The results are presented in Figure~\ref{fig:h2_gto} for both HF and DFT.

%\ML{I think throughout you need to be more careful in clarifying what you are using as a provisional basis vs. comparison against literal GTO. There are many places below. Right now it is too hard to tell and will confuse people. Golden rule = minimize reader frustration} For the GTO-only case, we employ the cc-pV$n$Z basis set centered on each hydrogen atom \ML{for n = blah, blah, ....}. The basis functions are orthogonalised and truncated such that each element contains a maximum of $N_\mathrm{trunc}=80$ functions. \ML{revisit notation. Comment why 80? The criterion  remains murky, looks suspicious} As a reference, ground-truth energies are computed using PySCF with the highly accurate cc-pV5Z basis set. The results are presented in Figure~\ref{fig:h2_gto} for both HF and DFT.
\begin{figure}[h]
    \centering
    \begin{tabular}{ |c||c|c|c||c|c|c|  }
     \hline
     \multicolumn{1}{|c||}{} & \multicolumn{3}{c||}{DG(HF)} & \multicolumn{3}{c|}{PySCF(HF)} \\
     \hline
     Basis & $N_\phi$ & Energy & Error & $N_\phi$ & Energy & Error \\
     \hline
     cc-pVDZ & 30 & -1.090161 & 0.001467 & 16 & -1.089673 & 0.001920 \\
     cc-pVTZ & 88 & -1.091264 & 0.000374 & 46 & -1.091166 & 0.000428 \\
     cc-pVQZ & 100 & -1.091529 & 0.000109 & 108 & -1.091481 & 0.000113 \\
     \hline
     \noalign{\vskip 8pt}
     \hline
     \multicolumn{1}{|c||}{} & \multicolumn{3}{c||}{DG(DFT)} & \multicolumn{3}{c|}{PySCF(DFT)} \\
     \hline
     Basis & $N_\phi$ & Energy & Error & $N_\phi$ & Energy & Error \\
     \hline
     cc-pVDZ & 33 & -1.106407 & 0.001494 & 16 & -1.106249 & 0.001651 \\
     cc-pVTZ & 88 & -1.107538 & 0.000363 & 46 & -1.107502 & 0.000399 \\
     cc-pVQZ & 100 & -1.107799 & 0.000102 & 108 & -1.107791 & 0.000110 \\
     \hline
    \end{tabular}
    \caption{HF and DFT energies for $\text{H}_2$ molecule using cc-pV$n$Z GTO basis sets. Reference energies are calculated using cc-pv5Z bases in PySCF. For the DG construction, the GTOs are used to construct a provisional basis while they are used directly as the computational basis in PySCF. In each case we report the number of basis functions $N_\phi$, as well as energies and energy errors relative to the reference energy.}
    \label{fig:h2_gto}
\end{figure}

For a given level of GTO approximation, slightly lower energies are obtained using the DG framework (where the GTOs are used to create a provisional DG basis), compared to the PySCF calculation (where the GTOs are used directly as continuous basis functions). This improvement can be attributed to the greater expressiveness of the discontinuous basis, as GTOs are duplicated and restricted in our framework, such that each provisional basis function has support only on a single element. The greater flexibility of the DG basis permits more effective representation of interstitial regions between atomic nuclei.

For DZ and TZ basis sets, we end up with a total number $N_\phi$ of DG basis functions that exceeds the size of the ordinary GTO basis. However, our DG basis enjoys improved sparsity of relevant basis integrals due to the disjoint supports of basis functions across elements. Strikingly, in the QZ case, our final DG basis set is actually smaller than the ordinary QZ basis set (and additionally still enjoys improved sparsity). 

%\ML{the upshots of your basis set choices are not clear. Want to interpret more intuitively. Are you using the same number of basis functions as vanilla GTO and getting better accuracy with better asymptotic scaling? Explain this}

Our next experiments consider a provisional basis built from both GTOs and polynomials. For this test, we use the smaller STO-3G Gaussian basis set, enriched with tensor product polynomial functions of degree $p=2,4,6$. The corresponding results are shown in Figure~\ref{fig:h2_gto_poly} for HF and DFT.

\begin{figure}[h]
    \centering
    \begin{tabular}{ |c||c|c|c|c|  }
     \hline
     \multicolumn{1}{|c||}{} & \multicolumn{2}{c|}{DG(HF)} & \multicolumn{2}{c|}{DG(DFT)} \\
     \hline
     p & Energy & Error & Energy & Error \\
     \hline
     - & -1.081895 & 0.009697 & -1.098372 & 0.009529 \\
     2 & -1.082367 & 0.009226 & -1.098670 & 0.009231 \\
     4 & -1.083886 & 0.007706 & -1.099606 & 0.008295 \\
     6 & -1.086619 & 0.004975 & -1.100011 & 0.007890 \\
     \hline
    \end{tabular}
    \caption{HF and DFT energies for $\mathrm{H}_2$ molecule using STO-3G + degree $p$ polynomial basis. For results in the the first row, no polynomials are added. Reference energies are calculated using cc-pv5Z bases in PySCF.}
    \label{fig:h2_gto_poly}
\end{figure}

In these experiments, chemical accuracy is not achieved when using STO-3G with added polynomials, even with the inclusion of high-degree polynomials on each element. This outcome is consistent with previous findings that highlight the limitations of polynomial basis sets in capturing sharp atomic cusps that remain insufficiently resolved here, given the use of a limited GTO basis.

\subsubsection{Mesh dependence} 
We investigate the dependence of the computed energy on the underlying DG mesh by performing a mesh-sensitivity experiment on the $\mathrm{H}_2$ molecule. The two hydrogen atoms are centered one unit apart $(0,0,0)$ and $(1,0,0)$.

The computational domain is partitioned into two DG elements (one associated with each atom), separated by an inter-element boundary located at position $x=h$. By varying $h \in [0,1]$, we slide the interface between the two atoms while keeping the nuclear positions fixed. For each choice of $h$, we construct the corresponding DG mesh and compute the HF and DFT energies using the cc-pVQZ basis as the provisional basis set.

A schematic of this setup and the resulting energies as a function of the interface position $h$ are reported in Figure~\ref{fig:h2_sliding}. We observe that the HF and DFT energies to vary only very slightly, on the order of $10^{-5}$, as $h$ is varied. This observation extends all the way to $h=0$ (or, by symmetry, $h=1$), where the interface sits on one of the atoms. As such, we do not expect our calculated energies to depend heavily on the initial mesh construction.

\begin{figure}[ht]
\centering
        
    \begin{minipage}[c]{0.49\textwidth}
    \centering
    \begin{tikzpicture}
    
      % Styles
      \tikzset{
        hydrogen/.style={circle, draw=black, fill=white, minimum size=8mm, inner sep=0pt},
        bond/.style={line width=1.2pt}
      }
    
      % Atom coordinates
      \coordinate (H1) at (-1,0);
      \coordinate (H2) at ( 1,0);
    
      % Bond
      \draw[bond] (H1) -- (H2);
    
      % Shortened brace (avoid circles)
      \draw[
        decorate,
        decoration={brace, amplitude=5pt},
        yshift=10cm
      ]
      ($(H1) + (0,0.6cm)$) -- ($(H2) + (0,0.6cm)$)
      node[midway, above=8pt] {$h \in [0,1]$};
    
      % Endpoint labels
      \node[below=10pt] at (H1) {$h=0$};
      \node[below=10pt] at (H2) {$h=1$};
    
      % Atoms
      \node[hydrogen] at (H1) {\textbf{H}};
      \node[hydrogen] at (H2) {\textbf{H}};
    
    \end{tikzpicture}
    \end{minipage}
    \hfill
    \begin{minipage}[c]{0.49\textwidth}
    \centering
    \begin{tikzpicture}
    
      % Styles
      \tikzset{
        hydrogen/.style={circle, draw=black, fill=white, minimum size=8mm, inner sep=0pt},
        bond/.style={line width=1.2pt},
        box/.style={draw=blue, thick}
      }
    
      % Atom coordinates
      \coordinate (H1) at (-1,0);
      \coordinate (H2) at ( 1,0);
    
      % Bond
      \draw[bond] (H1) -- (H2);
    
      % Atoms
      \node[hydrogen] at (H1) {\textbf{H}};
      \node[hydrogen] at (H2) {\textbf{H}};
    
      % Bounding box limits
      \def\xmin{-2}
      \def\xmax{2}
      \def\ymin{-1}
      \def\ymax{1}
    
      % Correct split location at h=0.25
      \pgfmathsetmacro{\splitx}{-1 + 2*0.25}
    
      % Draw aligned boxes
      \draw[box] (\xmin,\ymin) rectangle (\splitx,\ymax);
      \draw[box] (\splitx,\ymin) rectangle (\xmax,\ymax);
    
      % Label h = 0.25 at bond location
      \path (H1) -- (H2) coordinate[pos=0.5] (Split);
      \node[below=30pt] at (Split) {$h=0.25$};
    
    \end{tikzpicture}
    \end{minipage}
    \vspace{5mm}
  
    \pgfplotsset{every tick label/.append style={font=\huge}}
    \begin{tikzpicture}[baseline=(current bounding box.center), % <----
        scale=0.5,transform shape]
          \begin{axis}[
                ylabel style={ yshift=2ex },
                yticklabel style = { /pgf/number format/.cd, fixed, precision=5 },
                ymin=-1.09158,
                ymax=-1.09145,
                ymajorgrids,
                xtick pos=left,
                ytick pos=left,
                xtick = {0.0, 0.25, 0.5},
                xlabel={$h$},
                xlabel style={font=\huge},
                xlabel style={yshift=-10pt},
                ylabel={Energy (HF)},
                ylabel style={font=\huge},
                ylabel style={yshift=60pt}
            ]
            \addplot[mark=none, color=black, mark size=3pt]
            table[ x=h, y=Hf ]{img/sliding.dat};
            
            \end{axis}
    \end{tikzpicture}
    \hspace{10mm}
    \begin{tikzpicture}[baseline=(current bounding box.center), % <----
        scale=0.5,transform shape]
          \begin{axis}[
                ylabel style={ yshift=2ex },
                yticklabel style = { /pgf/number format/.cd, fixed, precision=5 },
                ymin=-1.107847,
                ymax=-1.107753,
                ymajorgrids,
                xtick pos=left,
                ytick pos=left,
                xtick = {0.0, 0.25, 0.5},
                xlabel={$h$},
                xlabel style={font=\huge},
                xlabel style={yshift=-10pt},
                ylabel={Energy (DFT)},
                ylabel style={font=\huge},
                ylabel style={yshift=60pt}
            ]
            \addplot[mark=none, color=black, mark size=3pt]
            table[ x=h, y=Dft ]{img/sliding.dat};
            
            \end{axis}
    \end{tikzpicture}
    \caption{Top: schematic for the mesh sensitivity test, where the inter-element interface is positioned at $x = h \in [0,1]$. Bottom: HF and DFT energies as a function of $h \in [0, 1/2]$, obtained using cc-pVQZ bas the provisional basis set.}
    \label{fig:h2_sliding}
\end{figure}

\subsection{$\boldsymbol{\mathrm{LiH}}$}
Next we consider the $\mathrm{LiH}$ molecule. The computational domain is split into two elements with each atom occupying one element. The exact setup is shown in Figure \ref{fig:lih_setup}. Following the procedure described above in Section \ref{sect:finding_ntrunc}, we choose a maximum of $N_\mathrm{filt} = 90$ basis functions for each element.

\begin{figure}[h]
  \centering

  % Left: TikZ diagram
  \begin{minipage}[c]{0.5\textwidth}
    \centering
    \begin{tikzpicture}

      % Styles
      \tikzset{
        hydrogen/.style={circle, draw=black, fill=white, minimum size=8mm, inner sep=0pt},
        lithium/.style={circle, draw=black, fill=gray!30, minimum size=10mm, inner sep=0pt},
        bond/.style={line width=1.2pt},
        box/.style={draw=blue, thick}
      }

      % Atom coordinates
      \coordinate (H) at (0, 0);
      \coordinate (Li) at (3.0235618, 0);

      % Bond
      \draw[bond] (H) -- (Li);

      % Atoms
      \node[hydrogen] (H_atom) at (H) {\textbf{H}};
      \node[lithium]  (Li_atom) at (Li) {\textbf{Li}};

      % Bounding box
      \def\xmin{-1}
      \def\xmax{4.2}
      \def\ymin{-1}
      \def\ymax{1}
      \pgfmathsetmacro{\boxwidth}{(\xmax - \xmin)/2}

      % Two split rectangles
      \draw[box] (\xmin, \ymin) rectangle ++(\boxwidth, {\ymax - \ymin});
      \draw[box] (\xmin+\boxwidth, \ymin) rectangle ++(\boxwidth, {\ymax - \ymin});

    \end{tikzpicture}
  \end{minipage}%
  % Right: Table
  \begin{minipage}[c]{0.45\textwidth}
    \centering
    \begin{tabular}{@{}ll@{}}
      \toprule
      Atom & Coordinates \\
      \midrule
      H  & (0.0, 0.0, 0.0) \\
      Li & (3.0235618, 0.0, 0.0) \\
      \bottomrule
    \end{tabular}
  \end{minipage}

  \caption{Mesh setup for $\text{LiH}$ molecule. Domain is split into two elements with atom positions in Bohr radii shown on the right.}
  \label{fig:lih_setup}
\end{figure}

In Figure \ref{fig:lih_gto}, we show results for both HF and DFT using cc-pV$n$Z GTO bases for $n=2,3,4$ (DZ, TZ, QZ). As in the hydrogen test case, the GTO basis is used to construct the provisional DG basis. We compare results to a standard calculation in the corresponding GTO basis using PySCF.

Again we observe that in the DZ and TZ cases, the induced DG basis is noticeably larger than the ordinary GTO basis, but this trend reverses for the larger QZ basis. We also examine the energies of the first three virtual orbitals obtained from HF, depicted in Figure \ref{fig:lih_virtuals}. For this experiment we have used the cc-pVQZ basis, both directly and as a means to construct a DG basis. As a reference, we compare against the virtuals obtained using the aug-cc-pVQZ GTO basis in PySCF.

\begin{figure}[h]
    \centering
    \begin{tabular}{ |c||c|c|c||c|c|c|  }
     \hline
     \multicolumn{1}{|c||}{} & \multicolumn{3}{c||}{DG(HF)} & \multicolumn{3}{c|}{PySCF(HF)} \\
     \hline
     Basis & $N_\phi$ & Energy & Error & $N_\phi$ & Energy & Error \\
     \hline
     cc-pVDZ & 73 & -7.986163 & 0.001164 & 43 & -7.985790 & 0.001547 \\
     cc-pVTZ & 150 & -7.987134 & 0.000193 & 94 & -7.986964 & 0.000373 \\
     cc-pVQZ & 180 & -7.987292 & 0.000035 & 198 & -7.987228 & 0.000109 \\
     \hline
     \noalign{\vskip 5pt}
     \hline
     \multicolumn{1}{|c||}{} & \multicolumn{3}{c||}{DG(DFT)} & \multicolumn{3}{c|}{PySCF(DFT)} \\
     \hline
     Basis & $N_\phi$ & Energy & Error & $N_\phi$ & Energy & Error \\
     \hline
     cc-pVDZ & 73 & -7.918440 & 0.001156 & 43 & -7.918077 & 0.001519 \\
     cc-pVTZ & 150 & -7.919393 & 0.000203 & 94 & -7.919217 & 0.000378 \\
     cc-pVQZ & 180 & -7.919563 & 0.000033 & 198 & -7.919493 & 0.000103 \\
     \hline
     \end{tabular}
    \caption{HF and DFT energies for $\text{LiH}$ molecule using cc-pV$n$Z GTO basis sets. Reference energies are calculated using cc-pv5Z bases in PySCF. For the DG construction, the GTOs are used to construct a provisional basis while they are used directly as the computational basis in PySCF. In each case we report the number of basis functions $N_\phi$, as well as energies and energy errors relative to the reference energy.}
    \label{fig:lih_gto}
\end{figure}

We observe that lower virtual energies are obtained using the discontinuous basis. Chemical accuracy obtained for the first virtual energy using the DG basis, which is not the case for the standard GTO basis. 

\begin{figure}[h]
\centering
\begin{minipage}{0.4\textwidth}
  \begin{tikzpicture}
    \begin{axis}[
      width=\linewidth, height=6.3cm,
      ymin=-0.01, ymax=0.035,
      axis x line=none,
      axis y line=left,
      ylabel={Energy},
      tick align=outside,
      line width=0.8pt,
    ]

    % DG top 3 levels including degeneracies
    \addplot[black, mark=none, thick] coordinates {(0.4,-0.006397) (0.6,-0.006397)};
    \addplot[black, mark=none, ultra thick] coordinates {(0.4,0.0257471) (0.6,0.0257471)};
    \addplot[black, mark=none, ultra thick] coordinates {(0.4,0.0257471) (0.6,0.0257471)};

    % PySCF top 3 levels including degeneracies
    \addplot[black, mark=none, thick] coordinates {(0.7,-0.003294) (0.9,-0.003294)};
    \addplot[black, mark=none, ultra thick] coordinates {(0.7,0.030198) (0.9,0.030198)};
    \addplot[black, mark=none, ultra thick] coordinates {(0.7,0.030198) (0.9,0.030198)};

    % Ref top 3 levels including degeneracies
    \addplot[black, mark=none, thick] coordinates {(1.0,-0.007578) (1.2,-0.007578)};
    \addplot[black, mark=none, ultra thick] coordinates {(1.0,0.013379) (1.2,0.013379)};
    \addplot[black, mark=none, ultra thick] coordinates {(1.0,0.013379) (1.2,0.013379)};

    \end{axis}

    % Add column labels as nodes
    \node at (0.85,-0.2) {\textbf{DG}};
    \node at (2.25,-0.2) {\textbf{PySCF}};
    \node at (3.55,-0.2) {\textbf{Ref}};
  \end{tikzpicture}
\end{minipage}%
\hspace{2em}
\begin{minipage}{0.4\textwidth}
  \centering
  \begin{tabular}{@{}lccc@{}}
    \toprule
    & \textbf{DG} & \textbf{PySCF} & \textbf{Ref} \\
    \midrule
    1 & -0.006401 & -0.003294 & -0.007578 \\
    2 & 0.025821 & 0.030198 & 0.013379 \\
    3 & 0.025821 & 0.030198 & 0.013379 \\
    \bottomrule
  \end{tabular}
\end{minipage}
\caption{Comparison of $\mathrm{LiH}$ virtual energies from DG and PySCF using cc-pVQZ GTO basis. For the DG construction, the GTOs are used to construct a provisional basis while they are used directly as the computational basis in PySCF. As a reference, we compare with the aug-cc-pVQZ basis in PySCF. We plot the first three virtual energies obtained in each case. The top lines on the left are bolded to indicate degeneracy.}
\label{fig:lih_virtuals}
\end{figure}

\subsection{$\boldsymbol{\mathrm{H}_2 \mathrm{O}}$}

Next we consider the water molecule. The problem geometry in shown in Figure \ref{fig:water_setup}, where the domain is split into three elements each of which contains one atom. Following the procedure described above in Section \ref{sect:finding_ntrunc}, we choose a maximum of $N_\mathrm{filt} = 98$ basis functions for each element.

In Figure \ref{fig:h2o_gto}, we show results for both HF and DFT using cc-pV$n$Z GTO bases for $n=2,3,4$ (DZ, TZ, QZ). As in previous examples, the GTO basis is used to construct the provisional DG basis. We compare results to a standard calculation in the corresponding GTO basis using PySCF.

\begin{figure}[h]
  \centering

  % Left: TikZ molecule
  \begin{minipage}[c]{0.5\textwidth}
    \centering
    \begin{tikzpicture}

      % Atom styles
      \tikzset{
        oxygen/.style={circle, draw=black, fill=red, minimum size=12mm, inner sep=0pt},
        hydrogen/.style={circle, draw=black, fill=white, minimum size=8mm, inner sep=0pt},
        bond/.style={line width=1.2pt},
        box/.style={draw=blue, thick}
      }

      % Atom coordinates (flattened)
      \coordinate (O)  at (0, 0);
      \coordinate (H1) at (-1.43052268, 1.10926924);
      \coordinate (H2) at ( 1.43052268, 1.10926924);

      % Bonds
      \draw[bond] (O) -- (H1);
      \draw[bond] (O) -- (H2);

      % Atoms
      \node[oxygen]   (O_atom)  at (O)  {\textbf{O}};
      \node[hydrogen] (H1_atom) at (H1) {\textbf{H}};
      \node[hydrogen] (H2_atom) at (H2) {\textbf{H}};

      % Bounding box
      \def\xmin{-2.2}
      \def\xmax{2.2}
      \def\ymin{-1.2}
      \def\ymax{2.2}
      \pgfmathsetmacro{\boxwidth}{(\xmax - \xmin)/3}

      % Three boxes
      \draw[box] (\xmin, \ymin) rectangle ++(\boxwidth, {\ymax - \ymin});
      \draw[box] (\xmin+\boxwidth, \ymin) rectangle ++(\boxwidth, {\ymax - \ymin});
      \draw[box] (\xmin+2*\boxwidth, \ymin) rectangle ++(\boxwidth, {\ymax - \ymin});

    \end{tikzpicture}
  \end{minipage}%
  % Right: Table
  \begin{minipage}[c]{0.45\textwidth}
    \centering
    \begin{tabular}{@{}lll@{}}
      \toprule
      Atom & Coordinates \\
      \midrule
      H & (1.43052268, 1.10926924, 0.0) \\
      O & (0.0, 0.0, 0.0) \\
      H & (-1.43052268, 1.10926924, 0.0) \\
      \bottomrule
    \end{tabular}
  \end{minipage}

  \caption{Mesh setup for $\text{H}_2\text{O}$ molecule. Domain is split into three elements with atom positions in Bohr radii shown on the right.}
  \label{fig:water_setup}
\end{figure}

We note that the computational basis size for this example remains somewhat larger than that of the ordinary GTO basis even in the QZ case. This might be attributable to the fact that we do not choose $N_{\mathrm{filt}}$ adaptively for each element.

\begin{figure}[h]
    \centering
    \begin{tabular}{ |c||c|c|c||c|c|c|  }
     \hline
     \multicolumn{1}{|c||}{} & \multicolumn{3}{c||}{DG(HF)} & \multicolumn{3}{c|}{PySCF(HF)} \\
     \hline
     Basis & $N_\phi$ & Energy & Error & $N_\phi$ & Energy & Error \\
     \hline
     cc-pVDZ & 105 & -76.035217 & 0.031826 & 51 & -76.030384 & 0.036659 \\
     cc-pVTZ & 220 & -76.058029 & 0.009014 & 114 & -76.057214 & 0.009829 \\
     cc-pVQZ & 294 &  -76.065387 & 0.001656 & 255 & -76.064823 & 0.002220 \\
     \hline
     \noalign{\vskip 5pt}
     \hline
     \multicolumn{1}{|c||}{} & \multicolumn{3}{c||}{DG(DFT)} & \multicolumn{3}{c|}{PySCF(DFT)} \\
     \hline
     Basis & $N_\phi$ & Energy & Error & $N_\phi$ & Energy & Error \\
     \hline
     cc-pVDZ & 105 & -75.874622 & 0.038040 & 51 & -75.867762 & 0.044899 \\
     cc-pVTZ & 220 & -75.900554 & 0.012107 & 114 & -75.910937 & 0.013143 \\
     cc-pVQZ & 294 & -75.910075 & 0.002586 & 255 & -75.909232 & 0.003429 \\
     \hline
    \end{tabular}
    \caption{HF and DFT energies for $\text{H}_2\text{O}$ molecule using cc-pV$n$Z GTO basis sets. Reference energies are calculated using cc-pv5Z bases in PySCF. For the DG construction, the GTOs are used to construct a provisional basis while they are used directly as the computational basis in PySCF. In each case we report the number of basis functions $N_\phi$, as well as energies and energy errors relative to the reference energy.}
    \label{fig:h2o_gto}
\end{figure}

\subsection{$\boldsymbol{\mathrm{C}_6 \mathrm{H}_6}$}

We perform a final test on the planar benzene $\mathrm{C}_6 \mathrm{H}_6$ molecule. The computational domain is divided into six elements, with each element containing exactly one carbon and one hydrogen atom, shown in Figure \ref{fig:benzene_setup}. Following the procedure described above in Section \ref{sect:finding_ntrunc}, we choose a maximum of $N_\mathrm{filt} = 190$ basis functions for each element.

In Figure \ref{fig:c6h6_gto}, we show results for both HF and DFT using cc-pV$n$Z GTO basis sets for $n=2,3,4$ (DZ, TZ, QZ). As in previous tests, the GTO basis is used to construct the provisional DG basis. We compare results to a standard calculation in the corresponding GTO basis using PySCF.

For lower levels of GTO approximation, the induced DG basis is smaller than the ordinary GTO basis set, but the DG basis is actually smaller in the QZ case. Likewise, the DG basis sets for a given level of GTO approximation yield more accurate energies.

\begin{figure}[h]
\centering

% Left: molecule in a centered box
\begin{minipage}[c]{0.35\textwidth}
\centering
\begin{tikzpicture}[scale=1]

  % Styles
  \tikzset{
    carbon/.style={circle, draw=black, fill=black!80, minimum size=7mm, inner sep=0pt},
    hydrogen/.style={circle, draw=black, fill=white, minimum size=5mm, inner sep=0pt},
    bond/.style={line width=0.8pt},
    box/.style={draw=blue, thick}
  }

  % Bounding box limits (width 5, height 6)
  \def\xmin{-2.5}
  \def\xmax{2.5}
  \def\ymin{-3.0}
  \def\ymax{3.0}

  % Draw bounding box
  \draw[box] (\xmin,\ymin) rectangle (\xmax,\ymax);

  % Number of divisions
  \def\nx{3}
  \def\ny{2}

  % Calculate box widths and heights
  \pgfmathsetmacro{\boxwidth}{(\xmax - \xmin)/\nx}
  \pgfmathsetmacro{\boxheight}{(\ymax - \ymin)/\ny}

  % Draw vertical dividing lines
  \foreach \i in {1,...,\numexpr\nx-1} {
    \pgfmathsetmacro{\xpos}{\xmin + \i*\boxwidth}
    \draw[box] (\xpos, \ymin) -- (\xpos, \ymax);
  }

  % Draw horizontal dividing line
  \foreach \j in {1,...,\numexpr\ny-1} {
    \pgfmathsetmacro{\ypos}{\ymin + \j*\boxheight}
    \draw[box] (\xmin, \ypos) -- (\xmax, \ypos);
  }

  % Benzene molecule atom coordinates (ring radius 1.3, hydrogens 1.8)
  % Center molecule at (0,0) inside bounding box width 5 x height 6

  \def\rc{1.5} % carbon radius
  \def\rh{2.2} % hydrogen radius

  % Coordinates of carbons and hydrogens
  \foreach \i in {1,...,6} {
    % Angles start at 90° and go around in 60° steps
    \coordinate (C\i) at ({90+60*(\i-1)}:\rc);
    \coordinate (H\i) at ({90+60*(\i-1)}:\rh);
  }

  % Draw bonds between carbons
  \foreach \i in {1,...,6} {
    \pgfmathtruncatemacro{\j}{mod(\i,6)+1}
    \draw[bond] (C\i) -- (C\j);
  }

  % Draw bonds carbon to hydrogen
  \foreach \i in {1,...,6} {
    \draw[bond] (C\i) -- (H\i);
  }

  % Draw atoms
  \foreach \i in {1,...,6} {
    \node[carbon]   at (C\i) {\textbf{C}};
    \node[hydrogen] at (H\i) {\textbf{H}};
  }

\end{tikzpicture}
\end{minipage}%
% Right: table without label column
\begin{minipage}[c]{0.48\textwidth}
\centering
\begin{tabular}{@{}ll@{}}
\toprule
Atom & Coordinates\\
\midrule
C & (0.0, 2.63955055, 0.0) \\
H & (0.0, 4.70596607, 0.0) \\
C & (-2.28575603, 1.31978473, 0.0) \\
H & (-4.05890495, 2.35299248, 0.0) \\
C & (-2.28575603, -1.31978473, 0.0) \\
H & (-4.05890495, -2.35299248, 0.0) \\
C & (0.0, -2.63955055, 0.0) \\
H & (0.0, -4.70596607, 0.0) \\
C & (2.28575603, -1.31978473, 0.0) \\
H & (4.05890495, -2.35299248, 0.0) \\
C & (2.28575603, 1.31978473, 0.0) \\
H & (4.05890495, 2.35299248, 0.0) \\
\bottomrule
\end{tabular}
\end{minipage}
\caption{Mesh setup for benzene molecule. The domain is split into six elements, each containing one $\text{C}$ and one $\text{H}$ atom. The atom positions in units of Bohr radius shown on the right.}
\label{fig:benzene_setup}
\end{figure}

\begin{figure}[h]
    \centering
    \begin{tabular}{ |c||c|c|c||c|c|c|  }
     \hline
     \multicolumn{1}{|c||}{} & \multicolumn{3}{c||}{DG(HF)} & \multicolumn{3}{c|}{PySCF(HF)} \\
     \hline
     Basis & $N_\phi$ & Energy & Error & $N_\phi$ & Energy & Error \\
     \hline
     cc-pVDZ & 476 & -230.738084 & 0.057864 & 258 & -230.721710 & 0.074238 \\
     cc-pVTZ & 966 & -230.784913 & 0.011035 & 552 & -230.778295 & 0.017653 \\
     cc-pVQZ & 1140 & -230.794650 & 0.001299 & 1188 & -230.792519 & 0.003429 \\
     \hline
     \noalign{\vskip 5pt}
     \hline
     \multicolumn{1}{|c||}{} & \multicolumn{3}{c||}{DG(DFT)} & \multicolumn{3}{c|}{PySCF(DFT)} \\
     \hline
     Basis & $N_\phi$ & Energy & Error & $N_\phi$ & Energy & Error \\
     \hline
     cc-pVDZ & 476 & -230.139441 & 0.062650 & 258 & -230.135675 & 0.066417 \\
     cc-pVTZ & 966 & -230.187508 & 0.014583 & 552 & -230.184766 & 0.017326 \\
     cc-pVQZ & 1140 & -230.199044 & 0.003047 & 1188 & -230.198799 & 0.003292 \\
     \hline
    \end{tabular}
    \caption{HF and DFT energies for $\text{C}_6\text{H}_6$ molecule using cc-pV$n$Z GTO basis sets. Reference energies are calculated using cc-pv5Z bases in PySCF. For the DG construction, the GTOs are used to construct a provisional basis while they are used directly as the computational basis in PySCF. In each case we report the number of basis functions $N_\phi$, as well as energies and energy errors relative to the reference energy.}
    \label{fig:c6h6_gto}
\end{figure}

%\ML{There is one additional experiment which would be interesting. Which is to show that we can use some ridiculuous AHG basis as the provisional basis without really increasing the cost and getting ultra high accuracy. Can use highly accurate benchmark from paper with Steve, or possibly just run directly with PySCF if tractable}

\section{Conclusion} \label{sect:conclusion}
In this work, we have introduced a framework for constructing adaptive discontinuous basis functions for electronic structure theory calculations. The Symmetric Interior Penalty Discontinuous Galerkin method is applied to handle the discontinuities to ensure that the discretisation is well-defined. 

We have considered the use of Gaussian-type orbital and polynomial basis functions in our framework. The discontinuous construction allows for arbitrary combinations of these functions to be employed as basis functions  on each element while retaining orthogonality. Furthermore, we apply an adaptive truncation procedure that allows us to control the size of the discontinuous basis sets.

The discontinuous basis sets are tested on Hartree-Fock and density functional theory calculations. The number of basis functions in the discontinuous construction is $O(MN_\mathrm{filt})$, where $N_\mathrm{filt}$ is a fixed, moderate number (typically around $100 \,  N_a$ where $N_a$ is the number of atoms contained in a single element) and $M$ denotes the number of elements. This cost is capped even as the underlying GTO basis used to construct the DG basis grows. By contrast, a direct calculation using GTOs as the computational basis becomes increasingly costly.

%\ML{make it clear what this means in terms of overall scaling in size extensive limit}

Interestingly, we find that combining polynomial functions with GTOs in the provisional basis offers no significant accuracy gain over using GTOs alone in the HF and DFT settings. Nonetheless, polynomials offer a systematic path toward achieving a complete basis, which GTOs alone do not provide. An important future direction is to explore the question of systematic convergence for post-HF methods, where basis set incompleteness errors are more pronounced.

The code for this project is available open-source on GitHub at
\[
\texttt{github.com/yllpan/dgSCF}
\]
We plan to maintain this code and update it as we conduct further investigations with this framework. 

\section*{Acknowledgments}
The authors are grateful to Sandeep Sharma, Lin Lin, Zhen Huang, and Xiao Liu for helpful discussions. M.L. was partially supported by
the U.S. Department of Energy, Office of Science,
Accelerated Research in Quantum Computing Centers, Quantum Utility through Advanced Computational Quantum Algorithms, grant no. DE-SC0025572,
the Applied Mathematics Program of the US Department of
Energy (DOE) Office of Advanced Scientific Computing Research under contract number DE-AC02-05CH11231, and a Sloan Research Fellowship. Y.P. and M.L. were partially supported by the Hellman Fellows Fund.

\newpage
%\bibliographystyle{plain}
%\bibliography{references}

\printbibliography

\newpage

\appendix
\part*{Appendices}
\section{Gaussian Fourier approximation} \label{appendix:fourier}

Consider an arbitrary Gaussian function
\begin{equation} \label{eq:gaussian_def}
    g(x) = \exp(-\alpha(x-x_0)).
\end{equation}
Without loss of generality, we set the shift $x_0=0$. We want to approximate this function using a Fourier cosine series
\begin{equation}
    g(x) \approx \sum_{k=1}^{N_F} a_k \cos \bigg( \frac{k\pi x}{L_\alpha} \bigg),
\end{equation}
where $L_\alpha$ is dependent on the width of the Gaussian $\alpha$.

It is sufficient for the Fourier approximation to hold only on an interval where $g(x) > \delta$, for some small tolerance $\delta$, as outside of this region we can simply set $g(x)=0$. Taking $\delta = 10^{-9}$, this means that we pick 
\begin{equation}
    L_\alpha = \sqrt{\frac{9}{\alpha} \log(10) }.
\end{equation}
The Fourier coefficients $a_k$ can be found by taking a Fourier transform of the Gaussian $g(x)$: 
\begin{equation}
    a_k = \frac{1}{L_\alpha} \sqrt{\frac{\pi}{\alpha}} \exp\bigg( \frac{-\pi^2 k^2}{4\alpha L_\alpha^2} \bigg).
\end{equation}
We find that this expansion yields a good approximation of the Gaussian $g(x)$ using only a small number $N_F$ of cosines, which can be taken independently of $\alpha$. In practice, setting $N_F=12$ yields uniform accuracy of around $10^{-10}$ over the interval $[-L_\alpha,L_\alpha]$.

\section{Gaussian-polynomial integrals}
\label{appendix:gaussian}

Our goal is to compute the one-dimensional integral
\begin{equation}
    I_\mathrm{gp} = \int_a^b g(x) p(x) ~dx,
\end{equation}
where $g(x)$ is a Gaussian function \eqref{eq:gaussian_def}, and $p(x)$ an arbitrary polynomial. To perform the integral we can apply the expansion in Appendix \ref{appendix:fourier} to obtain
\begin{equation}
    I_\mathrm{gp} \approx \sum_{k=1}^{N_F} a_k \int_a^b \cos \bigg( \frac{k \pi x}{L_\alpha} \bigg) p(x) ~dx.
\end{equation}
This reduces the integral to a sum of one-dimensional cosine-polynomial product integrals. We describe how to compute such integrals stably in the next section of the appendix.

\section{Cosine-product integrals}
\label{appendix:cosine}

Now we describe how to compute one-dimensional integrals of the general form
\begin{equation}
    \int_a^b \cos(\beta x) f(x) ~dx,
\end{equation}
where $f(x)$ is either a Gaussian function \eqref{eq:gaussian_def} or a polynomial. 

In the first case, where $f(x)=g(x)$ is Gaussian, we again apply the trick of expanding the Gaussian as a cosine expansion
\begin{equation}
    \int_a^b \cos(\beta x) g(x) ~dx \approx \sum_{k=1}^{N_F} a_k \int_a^b \cos(\beta x) \cos \bigg( \frac{k \pi x}{L_\alpha} \bigg) ~dx.
\end{equation}
This reduces in general to the integral of a product of cosines which can be handled via standard trigonometric identities 
\begin{equation}
    \int \cos(\beta x) \cos( \gamma x ) ~dx = \begin{cases}
        ~\dfrac{1}{2}\!\left[\dfrac{\sin((\beta-\gamma)x)}{\beta-\gamma} + \dfrac{\sin((\beta+\gamma)x)}{\beta+\gamma}\right] + C, & \beta \ne \gamma, \\
        ~\dfrac{x}{2} + \dfrac{\sin(2\beta x)}{4\beta} + C, & \beta = \gamma,
    \end{cases}
\end{equation}
where $C$ is an integration constant.

In the second case, where $f(x) = p(x)$ is a polynomial, we first represent the function in a Legendre polynomial basis
\begin{equation}
    p(x) = \sum_{k=0}^q c_k P_k(x),
\end{equation}
where $q$ is the degree of the polynomial and $P_k(x)$ are the Legendre polynomials. This representation improves numerical stability in practice for high polynomial powers $q$.

We then compute the integrals 
\begin{equation}
    I_k = \int_{-1}^{1} \cos( \beta x) P_k(x) ~dx, \quad J_k = \int_{-1}^{1} \sin( \beta x) P_k(x) ~dx, \quad k = 0, \ldots, q
\end{equation}
via recurrence relations. For general intervals, the integrals can be computed via a simple change of variables.

To proceed, recall that the Legendre polynomials satisfy the well-known recurrence
\begin{equation}
    P_k(x) = \frac{1}{2k+1} \frac{d}{dx} \bigg[ P_{k+1}(x) - P_{k-1}(x) \bigg],
\end{equation}
which allows us to compute the relations
\begin{equation}
    \begin{split}
        I_k &= \frac{1}{2k+1} \bigg[ \bigg( P_{k+1}(x) - P_{k-1}(x) \bigg) \cos(\beta x) \bigg]_a^b - \frac{\beta}{2k-1} \bigg[ J_{k+1} - J_{k-1} \bigg], \\
        J_k &= \frac{1}{2k+1} \bigg[ \bigg( P_{k+1}(x) - P_{k-1}(x) \bigg) \sin(\beta x) \bigg]_a^b + \frac{\beta}{2k-1} \bigg[ I_{k+1} - I_{k-1} \bigg],
    \end{split}
\end{equation}
which allows us to inductively construct the sequences $I_k, J_k$ ($k=0,1,2,\ldots$) by building from the base cases 
\begin{equation}
    I_0 = \frac{2\sin(\beta)}{\beta}, \quad I_1 = 0, \quad
    J_0 = 0, \quad J_1 = \frac{2\sin(\beta) -2\beta\cos(\beta)}{\beta^2}.
\end{equation}

\end{document}